\documentclass[iop,twocolappendix]{emulateapj}
\usepackage{enumerate}
\usepackage{amsmath}

\usepackage{multirow}
\usepackage{bigstrut}

\usepackage{verbatim} 
\usepackage{color} 
\usepackage[usenames,dvipsnames]{xcolor}

\shorttitle{Effects of Hot Gas Halo in Distant Encounters}
\shortauthors{Hwang \& Park}

\begin{document}
\title{Effects of hot halo gas on the star formation and mass transfer during distant galaxy$-$galaxy encounters}
\author{Jeong-Sun Hwang \altaffilmark{1, 2} \& 
Changbom Park \altaffilmark{1}}
\affil{$^1$ School of Physics, Korea Institute for Advanced Study, 85 Hoegi-ro, Dongdaemun-gu, Seoul 130-722, Korea; 
jshwang@kias.re.kr, cbp@kias.re.kr}
\affil{$^2$ School of Space Research, Kyung Hee University, Yongin, Gyeonggi-do 446-701, Korea}

\begin{abstract}
We use $N$-body/smoothed particle hydrodynamics simulations of encounters 
between an early-type galaxy (ETG)
and a late-type galaxy (LTG) 
to study the effects of hot halo gas 
on the evolution 
for a case 
with the mass ratio of 
the ETG to LTG 
of 2:1 and 
the closest approach distance of $\sim$100~kpc. 
We find that the dynamics of the cold disk gas 
in the tidal bridge and the amount of the newly formed stars 
depend strongly on 
the existence of a gas halo. 
In the run of interacting galaxies 
not having a hot gas halo, 
the gas and stars accreted into the ETG
do not include newly formed stars.   
However, in the run using the ETG with a gas halo 
and the LTG without a gas halo, 
a shock forms along the disk gas tidal bridge and 
induces star formation near the closest approach. 
The shock front is parallel to a channel along which the cold gas 
flows toward the center of the ETG. 
As a result, the ETG can accrete star-forming cold gas and 
newly born stars at and near its center. 
When both galaxies have hot gas halos,   
a shock is formed between the two gas halos 
somewhat before the closest approach. 
The shock hinders the growth of the cold gas bridge to 
the ETG and also ionizes it. 
Only some of the disk stars transfer through the stellar bridge. 
We conclude that the hot halo gas 
can give significant hydrodynamic effects 
during distant encounters.
\end{abstract}



\section{INTRODUCTION}

Interacting galaxy pairs have been actively studied 
both observationally and numerically over the past few decades 
(\citealt{Toomre_Toomre1972}; \citealt{Barnes_Hernquist1992}; 
\citealt{Struck2006}; \citealt{Smith+2007}; among many others). 
While the effects of galaxy mergers 
have been more intensely studied 
(e.g., \citealt{Springel+2005}; \citealt{Cox+2008}; 
\citealt{Genel+2008}; \citealt{Naab+2009}), 
the interest in a whole variety of interactions, 
including flybys and distant encounters 
(before or without mergers), 
is growing only recently 
(e.g., \citealt{Park+2008};  
\citealt{Hwang_Park2009}; 
\citealt{Park_Choi2009}; 
\citealt{Park_Hwang2009}; 
\citealt{Sinha2012}; 
\citealt{L'Huillier+2015}). 
For example, \citet{Park+2008} presented a unified picture 
on the evolution of galaxy morphology and luminosity 
through distant interactions 
and mergers by analyzing the Sloan Digital Sky Survey (SDSS;
\citealt{York+2000}) samples.
In particular, they found the strong dependence of galaxy morphology on 
the local environment set up by the nearest neighbor galaxy. 
It was claimed that hydrodynamic effects of cold gas as well as 
hot diffuse halo gas play an important role in the evolution of 
galaxy color and star formation activity. 
It was also shown that galaxy$-$galaxy encounters 
are very important in a cluster environment in star formation quenching 
through hydrodynamic interactions of late-type galaxies (LTGs) 
with cluster early-type galaxies (ETGs) (\citealt{Park_Hwang2009}).

So far, however, the overall effects of hot halo gas 
on galaxy evolution during various interactions 
have not been well understood.     
They can be tackled with the use of hydrodynamical models 
that include both cold and hot gas.  
Recently, \citet{Moster+2011} studied the role of hot gas 
in major mergers of disk galaxies, 
including for the first time a gradually cooling hot gas halo 
in the galactic-scale simulations. 
They showed that the hot gas component affects 
the star formation rates (SFRs), the efficiency of the starburst, 
and the kinematics and internal structure of the merger remnants.
In a subsequent study (\citealt{Moster+2012}), 
they also examined the effects of hot gas on disk thickening 
in minor mergers. 
They found that a cooling hot halo acts to reduce thickening 
by re-forming a massive thin disk in the post-merger. 
These studies have addressed that the results of interactions can be 
significantly affected by a hot gas halo component.

In order to further investigate the effects of hot halo gas 
in a wide range of interactions, 
we have been performing a series of 
$N$-body/hydrodynamic simulations, including 
a cooling diffuse hot gas halo in our galaxy models.
We presented the results of the first set of simulations 
in \citet{Hwang+2013} (hereafter Paper~I). 
In that work, we tested different density profiles for the gas halo 
with varying mass and initial spin in our Milky Way Galaxy$-$like models  
and examined how the adopted gas halos create differences 
in the evolution of the isolated galaxies, 
particularly in the gas dissipation and 
the star formation activity in the disk.

In the current work, 
we perform a set of simulations of 
interactions between an ETG and an LTG.
This work is motivated by the observational study 
by \citet{Park+2008}, 
where some interesting examples of ETGs 
having a late-type neighbor were presented, 
such as a blue elliptical-looking galaxy 
with strong star formation at its center, 
and a red galaxy having spiral-looking arms 
(see Figure~4 of their paper).   
In this work we aim to provide a 
set of 
simulated interacting galaxy systems that 
can be compared with observations in the follow-up studies.
Our study focuses on, for the first time, 
the effects of hot halo gas 
during distant encounters  
on the changes of galactic properties.
Specifically, we intend to investigate 
whether and how the presence of a hot gas halo component 
can cause differences in the star formation activities, 
the development and location of the tidal bridge, 
and the amount and dynamics of 
the material transferring through
the bridge 
to the ETG, 
as in the examples mentioned above, 
with a fixed set of orbital parameters.

We will continue to study
more cases of various interactions 
in different environments, emphasizing the role of hot halo gas 
in follow-up studies. 
For example, 
we plan to work on galaxy interactions  
in a rich galaxy cluster environment 
and investigate star formation quenching 
and the morphology transformation of LTGs 
influenced by the hot halo gas of neighboring ETGs, 
as well as the hot cluster gas.

This paper is organized as follows.   
We describe our initial galaxy models and the model parameters in Section~2, 
and the simulation code used for this work in Section~3. 
In Section~4, we present our simulation results of the distant encounters, 
explaining the orbital parameters, the evolution, 
and the star formation activities.
Finally, we summarize and discuss our results in Section~5.


\section{INITIAL GALAXY MODELS}

We use four different galaxy models for this work, among which 
two are LTG models~L and LH,  
and the other two are 
ETG models~E and EH 
(Table~1). 
We generate the models by using the 
ZENO software package (\citealp{Barnes2011}).
ZENO allows one to construct 
multiple spherical and disk components 
of a galaxy system in mutual equilibrium 
with user-specified density profiles 
in collisionless or gaseous form.

\begin{deluxetable*}{llcccc}
\centering
\tablecolumns{6}
\tablewidth{0pc}
\tablecaption{Initial galaxy models \label{tab02}}
\tablehead{
\colhead{Model components} &
\colhead{Model parameters} &
\colhead{Model L} &
\colhead{Model LH}&
\colhead{Model E} &
\colhead{Model EH}
}
\startdata
\\
& $M_{\rm tot}$\tablenotemark{a} ($10^{10}$ M$_{\odot}$)
       & 126 &  126 & 252  & 252\\
& $f_{\rm dg}$ (disk gas fraction)\tablenotemark{b}
    & 0.12 & 0.12  & $\cdots$  & $\cdots$  \\
& $f_{\rm hg}$ (halo gas fraction)\tablenotemark{c}
    & $\cdots$  & 0.01  & $\cdots$  & 0.01 \\
\hline
\\
\bf{Star disk:}\\ 
Disk model &   & Exponential  & Exponential & $\cdots$ & $\cdots$  \\
$a_{\rm ds}$ (kpc) & Star disk scale length & 3.5  &  3.5 & $\cdots$ & $\cdots$ \\
$z_{\rm ds}$ (kpc)            & Vertical disk scale height    & 0.35&  0.35 & $\cdots$ & $\cdots$\\
$b_{\rm ds}$ (kpc) & Outer disk cutoff radius & 42 &   42 & $\cdots$  & $\cdots$ \\
$M_{\rm ds}$ ($10^{10}$ M$_{\odot}$)  & Total star disk mass   & 4.4  &   4.4 & $\cdots$& $\cdots$ \\
$N_{\rm ds}$       & Number of particles  &  204 800   &  204 800 & $\cdots$ & $\cdots$ \\
$m_{\rm ds}$ ($10^{5}$ M$_{\odot}$) & Mass of individual particles
                                          & 2.15
                                          & 2.15
                                          & $\cdots$
                                          & $\cdots$
                                               \\
$\epsilon_{\rm ds}$ (kpc) & Gravitational softening length  &
                             0.07 & 0.07 & $\cdots$ & $\cdots$ \\
\hline
\\
\bf{Gas disk:}\\ 
Disk model &   &  Exponential  & Exponential  & $\cdots$ &   $\cdots$ \\
$a_{\rm dg}$ (kpc) & Gas disk scale length      & 8.75
                                                   & 8.75
                                                   &   $\cdots$  &   $\cdots$ \\
$z_{\rm dg}$ (kpc) & Vertical disk scale height    & 0.35  &    0.35 &   $\cdots$  &   $\cdots$ \\
$b_{\rm dg}$ (kpc) & Outer disk cutoff radius & 105 &   105 &   $\cdots$  &   $\cdots$ \\
$M_{\rm dg}$ ($10^{10}$ M$_{\odot}$) & Total gas disk mass   & 0.6   &   0.6 &   $\cdots$ &   $\cdots$ \\
$N_{\rm dg}$  & Number of particles   & 81 920 &    81 920 &   $\cdots$   &   $\cdots$ \\
$m_{\rm dg}$ ($10^{5}$ M$_{\odot}$) & Mass of individual particles
                                              & 0.73
                                              & 0.73
                                              & $\cdots$
                                              & $\cdots$
                                              \\
$\epsilon_{\rm dg}$ (kpc) & Gravitational softening length  &
                             0.04 & 0.04  &  $\cdots$ & $\cdots$ \\
\hline
\\
\bf{DM halo:} \\
Halo model & & NFW & NFW & NFW &   NFW\\
$a_{\rm hd}$ (kpc)        & DM halo scale length  &   21   &   21    & 21   &   21  \\
$b_{\rm hd}$ (kpc)        & Tapering radius  & 84 & 84   & 84 &  84   \\
$M_{\rm hd}(a_{\rm hd})$ ($10^{10}$ M$_{\odot}$) & Mass within radius $a_{\rm hd}$
                                                  & 12.35  & 12.23 & 24.71 & 24.46       \\
$M_{\rm hd}(\infty)=M_{\rm hd}$ ($10^{10}$ M$_{\odot}$) & Total DM halo mass  &
        120   & 118.8 & 240 & 237.6 \\
$N_{\rm hd}$ & Number of particles
          & 1 024 000  &  1 024 000 &  2 048 000 &  2 048 000\\
$m_{\rm hd}$ ($10^{5}$ M$_{\odot}$) & Mass of individual particles
                                            & 11.7 &
                                              11.6 &
                                              11.7 &
                                              11.6  \\
$\epsilon_{\rm hd}$ (kpc)  & Gravitational softening length  &
                             0.16 & 0.16 & 0.16 & 0.16   \\
\hline
\\
\bf{Gas halo:} \\
Halo model &   &  $\cdots$   & Isothermal &  $\cdots$ &  Isothermal\\
$a_{\rm hg}$ (kpc)
              & Core radius &  $\cdots$   &  10.5  &    $\cdots$   &  10.5\\
$b_{\rm hg}$ (kpc)
             & Tapering radius   &  $\cdots$ & 420 &  $\cdots$  & 420\\
$M_{\rm hg}$ ($10^{10}$ M$_{\odot}$) & Total gas halo mass
		&  $\cdots$   & 1.2  &   $\cdots$ & 2.4 \\
$N_{\rm hg}$ & Number of particles
        &  $\cdots$   & 163 840  &   $\cdots$  & 327 680 \\
$m_{\rm hg}$  ($10^{5}$ M$_{\odot}$) & Mass of individual particles
                                            &  $\cdots$
                                            &  0.73
                                            & $\cdots$
                                            & 0.73
                                            \\
$\epsilon_{\rm hg}$ (kpc)  & Gravitational softening length  &
                              $\cdots$  & 0.04  &  $\cdots$ & 0.04  \\
\hline
\\
\bf{Bulge:} \\
Bulge model & &  Hernquist &  Hernquist &      Hernquist & Hernquist\\
$a_{\rm b}$ (kpc)  & Bulge scale length   & 0.7 &  0.7 & 2.8 & 2.8\\
$b_{\rm b}$ (kpc) & Truncation radius  & 140 & 140 &  280 &  280 \\
$M_{\rm b}$ ($10^{10}$ M$_{\odot}$)  & Total bulge mass   & 1  &  1 & 12  & 12\\
$N_{\rm b}$ & Number of particles
          & 45 056   & 45 056 & 540 672 &  540 672 \\
$m_{\rm b}$ ($10^{5}$M$_{\odot}$) & Mass of individual particles
                                    & 2.22
                                    & 2.22
                                    & 2.22
                                    & 2.22
                                    \\
$\epsilon_{\rm b}$ (kpc) & Gravitational softening length  &
                               0.07 & 0.07 & 0.07  & 0.07                    
\enddata
\tablenotetext{a}{$M_{\rm tot} = M_{\rm d} + M_{\rm h} + M_{\rm b} 
                               =  M_{\rm ds} + M_{\rm dg} +  M_{\rm hd} + M_{\rm hg}  + M_{\rm b}$ }
\tablenotetext{b}{$f_{\rm dg} = M_{\rm dg}/M_{\rm d}$}
\tablenotetext{c}{$f_{\rm hg} = M_{\rm hg}/M_{\rm h}$}
\end{deluxetable*}

\renewcommand{\multirowsetup}{\centering} 
\begin{table*}[t]
\caption{Three runs of distant encounters}
\vspace*{-2mm}
\begin{center}
\begin{tabular}{l|c c|c c | c c c}
\hline \hline
\multirow{3}{1.5cm}{ETG-LTG}& \multicolumn{2}{p{3.5cm}|}{\centering ETG}  & \multicolumn{2}{p{3.5cm}|}{\centering LTG} & 
\multicolumn{3}{p{4cm}}{\centering 1st closest approach} \\
& \multicolumn{1}{c}{$x_{0},y_{0},z_{0}$} 
& \multicolumn{1}{c|}{$v_{x0},v_{y0},v_{z0}$} 
& \multicolumn{1}{c}{$x_{0},y_{0},z_{0}$} 
& \multicolumn{1}{c|}{$v_{x0},v_{y0},v_{z0}$} 
& \multicolumn{1}{c}{$dr$} 
& \multicolumn{1}{c}{$dv$} 
& \multicolumn{1}{c}{$t$}  
\\ 
& \multicolumn{1}{c} {(kpc)} 
& \multicolumn{1}{c|}{(km s$^{-1}$)} 
& \multicolumn{1}{c} {(kpc)} 
& \multicolumn{1}{c|}{(km s$^{-1}$)} 
& \multicolumn{1}{c} {(kpc)} 
& \multicolumn{1}{c} {(km s$^{-1}$)} 
& \multicolumn{1}{c} {(Gyr)}  
\\      
\hline
E-L    & 0, 0, 0   & 0, 0, 0  & $-$1200, 265, 0 & 200, 0, 0  &  94.3 &  495.6 & 4.75 \\
EH-L   & 0, 0, 0   & 0, 0, 0  & $-$1200, 265, 0 & 200, 0, 0  &  94.4 &  494.8 & 4.75 \\
EH-LH  & 0, 0, 0   & 0, 0, 0  & $-$1200, 265, 0 & 200, 0, 0  &  94.8 &  495.5 & 4.75 \\
\hline
\end{tabular}
\end{center}
\end{table*}


\subsection{Density Profiles}

We first present the specific form of the density profile 
adopted for each component of our galaxy models 
(refer to \citealt{Barnes_Hibbard2009} for more details; see also Paper~I).

Both star and gas disks in the LTG models 
follow an exponential surface density profile 
and a sech$^2$ vertical profile: 
\begin{equation}
  \rho_{\rm dc}(R,z) =
    \frac{M_{\rm dc}}{4 \pi a_{\rm dc}^2 z_{\rm dc}} \,
      e^{- R / a_{\rm dc}} \,
        \mathrm{sech}^2 \left( \frac{\it z}{\it z_{\rm dc}} \right) \, ,
  \label{eq1}
\end{equation}
where the subscript ``$\rm c$" stands for either ``$\rm s$" 
for the star disk component or ``$\rm g$" for the gas disk component; 
$R$ represents the cylindrical radius $(x^2+y^2)^{1/2}$; 
$M_{\rm dc}$ is the mass of the star or gas disk; 
$a_{\rm dc}$ and $z_{\rm dc}$ are the radial scale length and 
the vertical scale height of the disk, respectively; and 
$a_{\rm dg}$ is chosen to be 
2.5 times larger than $a_{\rm ds}$ (Table~1).

The dark matter (DM) halo in all of our models has 
has a Navarro$-$Frenk$-$White (NFW) 
profile (\citealt{Navarro+1996}). 
Because the cumulative mass of the profile diverges at large radii,
we apply an exponential taper at radii larger than $b_{\rm hd}$:
\begin{equation}
\rho_{\rm hd}(r) =
  \left\{
  \begin{array}{ll}
   \displaystyle
   \frac{M_{\rm hd}( a_{\rm hd})}{4 \pi (\ln(2) - \frac{1}{2})} \,
   \frac{1}{r (r + a_{\rm hd})^2} &
	    {\rm for} \,\, r \le b_{\rm hd} \, , \\ [0.4cm]
   \displaystyle
   \rho_{\rm hd}^{*} \, \left(\frac{b_{\rm hd}}{r}\right)^2 \,
	    e^{-2 \beta (r / b_{\rm hd} -1)} &
	    {\rm for} \,\, r > b_{\rm hd} \, , \\
  \end{array}
  \right.
\label{eq2}
\end{equation}
where $a_{\rm hd}$ is the radial scale length of the DM halo, 
and $M_{\rm hd}(a_{\rm hd})$ is the mass within $a_{\rm hd}$.

Two of our models (LH and EH) possess 
a nonsingular isothermal gas halo with 
a taper:
\begin{equation}
  \rho_{\rm hg}(r) =
	\displaystyle
          \frac{f_{\rm norm}  M_{\rm hg}}{2 {\pi} \sqrt{\pi} \, b_{\rm hg}} \,
          \frac{1}{r^2 + a_{\rm hg}^2} \,  e^{- (r / b_{\rm hg})^2}
	       \, ,  \\
  \label{eq3}
\end{equation}
where $a_{\rm hg}$ is the radius of the inner core 
with a constant density, 
$b_{\rm hg}$ is the radius of taper, 
and $M_{\rm hg}$ is the mass of the gas halo.
The isothermal profile is chosen because it is similar to 
the $\beta$~profile that is commonly used 
to fit the hot gas observed in galaxy clusters.

The mass density of the bulge in all models follows
a \citet{Hernquist1990} profile with 
truncation at large radii from $b_{\rm b}$ (Table~1): 
\begin{equation}
  \rho_{\rm b}(r) =
    \left\{
      \begin{array}{ll}
	\displaystyle
          \frac{a_{\rm b} M_{\rm b}}{2 \pi} \,
	      \frac{1}{r (a_{\rm b} + r)^{3}} \, &
	     {\rm for} \,\, r \le b_{\rm b} \, , \\ [0.4cm]
	\displaystyle
	  \rho_{\rm b}^{*} \, \left(\frac{b_{\rm b}}{r}\right)^2 \,
	      e^{-2 r / b_{\rm b}} \,  &
	     {\rm for} \,\, r > b_{\rm b} \, , \\
      \end{array}
    \right.
  \label{eq4}
\end{equation}
where $a_{\rm b}$ is the radial scale length of the bulge; 
$M_{\rm b}$ is the bulge mass. 
The bulge consists of stars only.

With the above density profiles, 
the $N$-body realizations of our galaxy models are made 
by using the ZENO programs.    
We have run each model in isolation 
to check the stability 
before performing simulations of distant encounters.
In Appendix~A, we present the evolution of model~LH, 
which possesses both a gas disk and a gas halo.


\subsection{LTG Models}

The total mass of each LTG model is set 
to $M_{\rm tot}=126 \times 10^{10}\,\rm{M_{\odot}}$ (Table~1), 
similar to that of the Milky Way Galaxy.

Model~L consists of the four components: stellar disk, gas disk, 
DM halo, and bulge. 
The mass of each component   
in units of $10^{10}\,\rm{M_{\odot}}$ is
$M_{\rm ds} = 4.4$, $M_{\rm dg} = 0.6$,
$M_{\rm hd} = 120$, and $M_{\rm b} = 1$, respectively (Table~1). 
The total disk mass is thus 
$M_{\rm d} = M_{\rm ds} + M_{\rm dg} = 5.0 \times 10^{10}\,\rm{M_{\odot}}$.
The gas fraction in the disk, 
defined as $f_{\rm dg} = M_{\rm dg}/M_{\rm d}$, 
is 0.12.
The number of star or gas particles 
distributed in each component is  
$N_{\rm ds} = 204,800$, $N_{\rm dg} = 81,920$,
$N_{\rm hd} = 1,024,000$, and $N_{\rm b} = 45,056$.

Model~LH has an additional component of a gaseous halo.
The gas halo component has 
$M_{\rm hg} = 1.2 \times 10^{10}\,\rm{M_{\odot}}$ 
and $N_{\rm hg} = 163,840$.
Because the DM halo mass is set to 
$M_{\rm hd} = 118.8 \times 10^{10}\,\rm{M_{\odot}}$, 
the total halo mass is 
$M_{\rm h} = 120 \times 10^{10}\,\rm{M_{\odot}}$, 
equal to that of model~L. 
The halo gas fraction is $f_{\rm hg} = M_{\rm hg}/M_{\rm h}$ = 0.01.
The gas halo is 
twice more massive than the gas disk in this model. 
The mass of a single gas particle (either in the disk or in the halo) 
is the same at the initial time.

Both star and gas disks in the LTG models 
have an exponential surface density profile 
and a sech$^2$ vertical profile (Equation~(1)). 
The radial and vertical scale lengths of the star disk 
are set to 
$a_{\rm ds} = 3.5$~kpc and $z_{\rm ds} = 0.35$~kpc, respectively.
Those of the gas disk are set to 
$a_{\rm dg} = 2.5 \times a_{\rm ds} = 8.75$~kpc 
and $z_{\rm dg} = z_{\rm ds}$ (\citealt{Moster+2011}). 
The mass distribution of the DM halo 
follows the NFW profile (Equation~(2)). 
Its radial scale is $a_{\rm hd} = 21$~kpc, 
which is adopted from \citet{McMillan_Dehnen2007}.
For the bulge, the Hernquist profile (Equation~(4)) is chosen 
with the length scale of 
$a_{\rm b} = 0.7$~kpc (\citealt{McMillan_Dehnen2007}). 
The gas halo included in model~LH follows 
the isothermal density profile (Equation~(3)) 
with the core radius of $a_{\rm hg} = 10.5$~kpc.

The disk gas particles in models~L and LH 
are set to rotate in the clockwise direction 
with the local circular velocities 
without any motion in the vertical direction 
at the initial time.
For the disk star particles in both models, 
velocity dispersions are added to the local circular velocities, 
as described 
in \citet{Barnes_Hibbard2009}.\footnote{ 
In Paper~I, the dispersions were added to 
the disk star particles in the models of type~DH, but were not 
applied in the models of type~D. 
}

The temperatures of the disk gas particles in both models 
are set to 
$T = 10,000$~K. 
For the halo gas particles in model~LH, 
the initial temperatures are determined 
in accordance with the hydrodynamic equilibrium of 
the halo gas.\footnote{The temperature distribution 
of the halo gas particles of model~LH is statistically 
the same as that of model~DHi in Paper~I.
Compared with the models in Paper~I, 
we use a larger number of particles for this work 
and accordingly adopt smaller gravitational softening lengths 
in all components. 
Except for the particle numbers and the dispersions of disk stars, 
all of the other key model parameters, such as scale length, 
truncation/tapering radius, 
and core radius, remain unchanged.}


\subsection{ETG Models}

The ETG models~E and EH do not have a disk component. 
The total mass of each ETG model 
is set to $252 \times 10^{10}\,\rm{M_{\odot}}$ (Table~1), 
twice more massive than that of the LTG models.

Model~E consists of two components, a DM halo with 
$M_{\rm hd} = 240 \times 10^{10}\,\rm{M_{\odot}}$ 
and a bulge with 
$M_{\rm b} = 12 \times 10^{10}\,\rm{M_{\odot}}$ (Table~1). 
The number of particles 
distributed in each component is    
$N_{\rm hd} = 2,048,000$ and $N_{\rm b} = 540,672$.
The masses of a particle in the DM halo and in the bulge of model~E 
are equal to those of model~L. 

Model~EH has a gas halo in addition to 
a DM halo and a bulge. 
The gas halo mass is 
$M_{\rm hg} = 2.4 \times 10^{10}\,\rm{M_{\odot}}$ 
with the halo gas fraction $f_{\rm hg}$ = 0.01.
The number of halo gas particles is $N_{\rm hg} = 327,680$.
The mass of the DM halo is 
$M_{\rm hd} = 237.6 \times 10^{10}\,\rm{M_{\odot}}$.
The total (DM + gas) halo mass and the bulge mass are 
equal to those of model~E.

In the ETG models, 
the DM halo and the bulge follow
the NFW profile (Equation~(2)) and the Hernquist profile (Equation~(4)), respectively, 
as in the LTG models. 
The scale length of the DM halo ($a_{\rm hd}$) 
and the core radius of the
gas halo ($a_{\rm hg}$) are equal to those of model LH (Table 1). 
The scale length of the bulge
component ($a_{\rm b}$) is set to 2.8~kpc.


\section{Simulation code and the parameters}

To perform simulations of interactions between the galaxy models,   
we use an early version of the 
GADGET-3 TreeSPH code (originally described in \citealt{Springel2005}). 
Gravitational force is calculated 
by means of the tree algorithm (\citealt{Barnes_Hut1986}), 
and the hydrodynamic force is computed 
with the smoothed particle hydrodynamics (SPH) method 
in the entropy conservative formulation \citep{Springel+2002}.
The code includes radiative cooling and heating 
for a primordial mixture of hydrogen and helium 
by photoionization (\citealt{Katz1996}).
Star formation and supernova feedback in the interstellar medium (ISM) 
are implemented using the effective multiphase model of 
\citet{Springel_Hernquist2003}.
A thermal instability operates for gas at high density 
over a threshold value of $\rho_{\rm th}$, 
and the ISM is treated as a statistical mixture of cold clouds 
and ambient hot medium. 
Star formation occurs in dense regions consuming the cold clouds. 
The consumption timescale is chosen to match 
the observations \citep{Kennicutt1998}.
Among the newly formed stars, the mass fraction of 
massive stars ($>8~\rm{M_{\odot}}$) 
is given by the parameter $\beta$.
The massive stars die instantly as supernovae 
releasing energy as heat to the ambient diffuse medium.  
Some cold clouds are also evaporated inside the supernova bubbles 
returning material to the ambient gas.
We do not include the galactic winds associated with star formation 
in this work.
We also do not consider 
active galactic nucleus feedbacks. 

For the parameters related to star formation and the feedbacks, 
we use standard values of the multiphase model in all of our simulations:
the star formation timescale is 
$t_{0}^{\star} = 2.1$~Gyr, and
the mass fraction of massive stars is $\beta = 0.1$.
The ``supernova temperature" is 
$T_{\rm SN} = 10^{8}$~K, and 
the temperature of cold clouds is $T_{\rm c} = 1000$~K.
The parameter value for supernova evaporation is $A_{0} = 1000$. 
In the code, 
the mass of a gas particle involving star formation 
is chosen to be divided in two: 
a new star particle has half of the original gas mass, 
and the mass of the gas particle decreases by half. 
Thus, 
the total number of particles in each model 
increases as star formation occurs.

We set the gravitational softening length for the DM halo particles 
to $\epsilon_{\rm hd} = 0.16$~kpc in all simulations. 
The value is about an order of magnitude smaller than 
the average distance between the DM particles in our models.
Then we set the maximum acceleration experienced by 
a particle ($f \propto m/\epsilon^2$, with the initial mass) 
to be equal in each component. 
For example, in the simulations of model~LH in isolation 
and in interaction with model~EH,  
the softening lengths 
for the gas (both halo gas and disk gas), 
disk star, and bulge particles are determined accordingly 
to be 0.04, 0.07, and 0.07~kpc, respectively.
The softening lengths for disk gas and disk star particles 
are comparable to the average distances between particles 
in the components.

To achieve better accuracy in the force computation,  
we chose the value of the parameter $\alpha$, which controls 
the accuracy of the relative cell-opening criteria 
(\citealt{Springel2005}), to be $0.0005$ in all simulations, 
which is about an order of magnitude smaller than 
the values popularly used in galactic-scale simulations.
We set the opening angle to $\theta = 0.3$, which is 
used for the first force computation.

\begin{figure*} [!hbt]
\centering
\includegraphics[width=15cm]{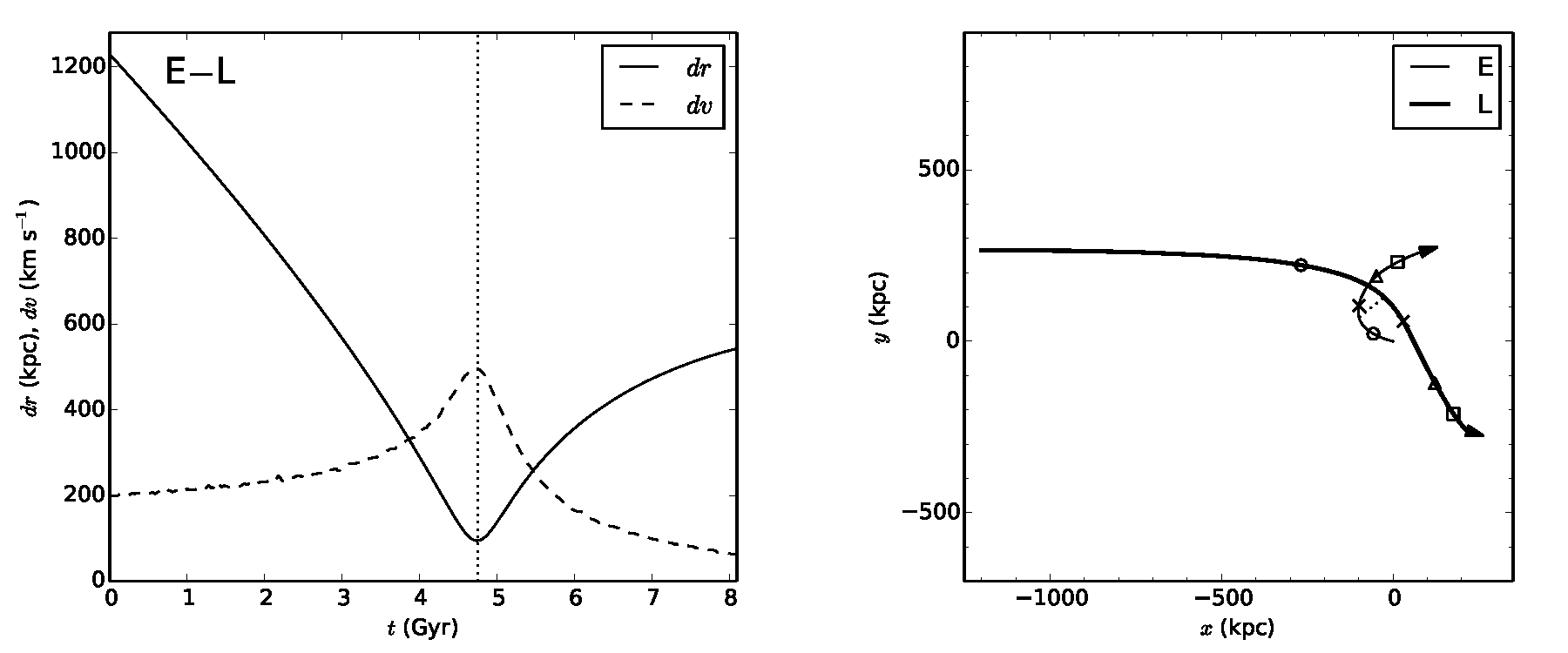}
\caption{{\it{Left}}: distance ($dr$, solid line) 
and velocity ($dv$, dashed line) 
of the late-type galaxy relative to the early-type galaxy 
with respect to time in run~E-L.
Here, $dr$ is measured between the center of mass positions of 
the bulges of the two galaxies, and 
$dv$ is the average velocity of the bulge particles of the late-type galaxy 
relative to that of the early-type galaxy. 
The closest approach between the two galaxies occurs  
at $t = 4.75$~Gyr and is marked with the vertical dotted line.
{\it {Right}}: 
orbital trajectories of the two galaxies in 
run~E-L seen in the $x$$-$$y$~plane.
The late-type galaxy L starts from the left 
going to the right (thick solid line) 
and the early-type galaxy E starts from the origin and moves upward 
(thin solid line), as indicated with the arrowheads.
The positions of the galaxies at $t =$ 4, 5, 6, and 7~Gyr
are marked with circles, crosses, triangles, and squares 
on the trajectories, respectively.
The closest approach between the galaxies 
is marked with the dotted line. 
}
\end{figure*}

\begin{figure*} [!hbt]
\centering
\includegraphics[width=13.5cm]{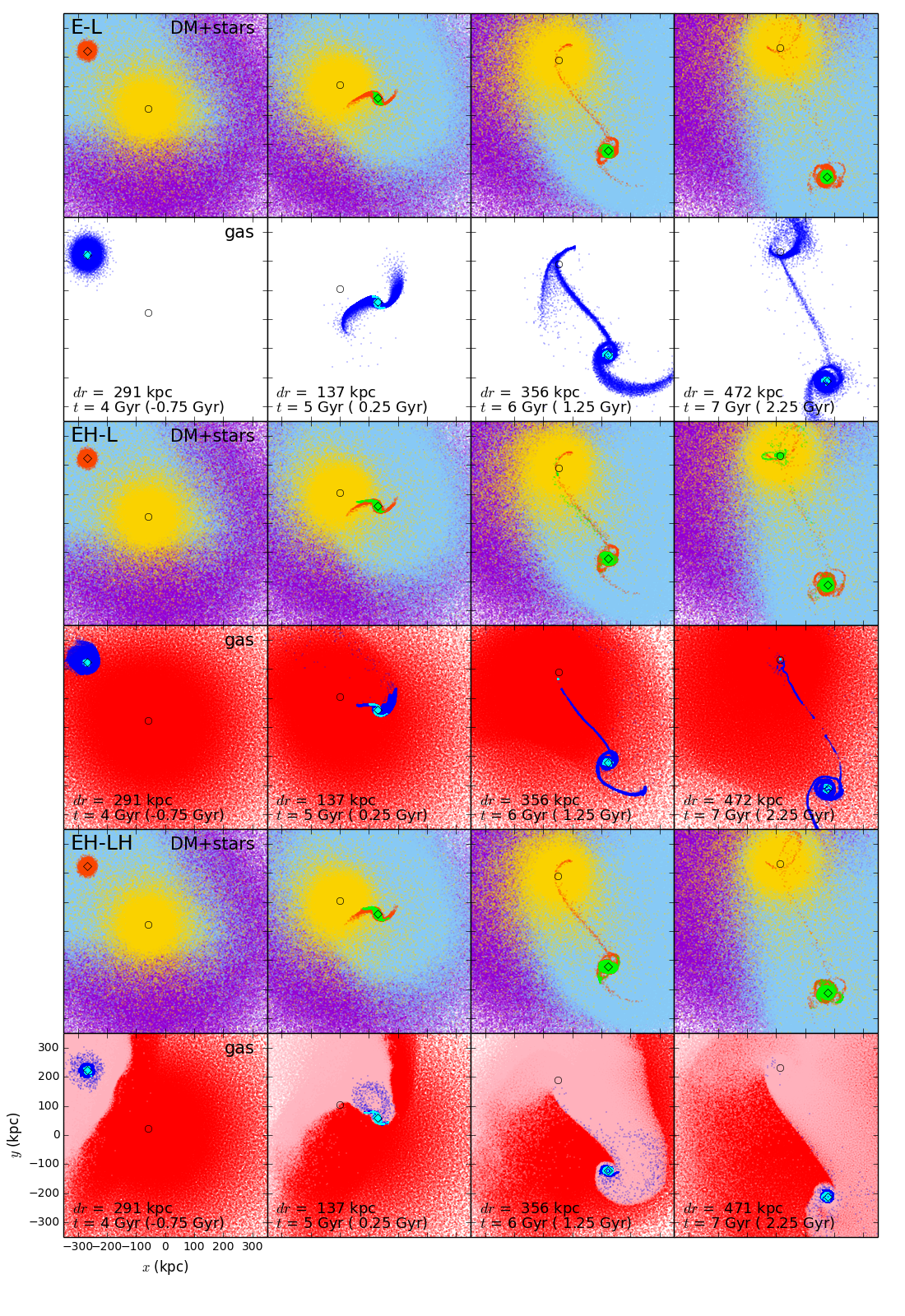}
\caption{
{\it{Top two rows}}: 
four snapshots of the distribution of the particles 
in run~E-L 
seen in the $x$$-$$y$ plane at $t$ = 4, 5, 6, and 7~Gyr 
(from the first to the fourth columns, respectively). 
The collisionless (DM and star) particles and 
the gas particles are displayed separately 
in the upper and the lower panels, respectively. 
In each panel, the positions of the center of mass of the bulge 
of the early- and the late-type galaxies are marked with 
a circle and a diamond, respectively. 
In the lower panels, 
the three-dimensional distance between the two bulges at each time 
is given; 
the time measured from the closest approach is also written 
in the parentheses. 
The different colors are used to distinguish the 
origin of the particles. In the upper panels, 
the violet and yellow points represent 
the particles initially set as the halo DM and the bulge stars 
of the early-type galaxy, respectively. 
The light-blue points denote the particles set as
the halo DM of the late-type galaxy. 
The orange points represent the ``old" stars 
initially set as the disk stars and those added onto the disk 
out of the gas originally set as the disk gas 
before the closest approach, i.e., $t < $4.75~Gyr. 
The green points represent the ``young" stars 
formed out of the initial disk gas 
since the closest approach, $t \geq$ 4.75~Gyr. 
(The bulge particles of the late-type galaxy 
are not displayed for simplicity.) 
In the lower panels, 
the blue points denote the non-star-forming gas particles, 
which were set as the disk gas of the late-type galaxy 
(and remain in gas). 
The star-forming gas particles 
having positive values of star formation rate 
at that instant are shown 
with the cyan points.   
{\it{Middle two rows}}: 
the same snapshots as the upper two rows, but from run~EH-L. 
The red points in the lower panels represent 
the particles originally set as the halo gas of the early-type galaxy EH.
{\it{Bottom two rows}}: 
the same snapshots as above, but from run~EH-LH. 
The pink points in the lower panels represent 
the particles originally set as the halo gas of the late-type galaxy LH.
The green points in the upper panels  
include (at the central part of the disk) 
a small amount of the young stars formed out of 
the gas originally set as the halo gas of LH.  
The cyan points in the lower panels also include 
a few star-forming gas particles that were initially 
set as the halo gas of LH.  
}
\end{figure*}

\begin{figure*} [!hbt]
\centering
\includegraphics[width=14cm]{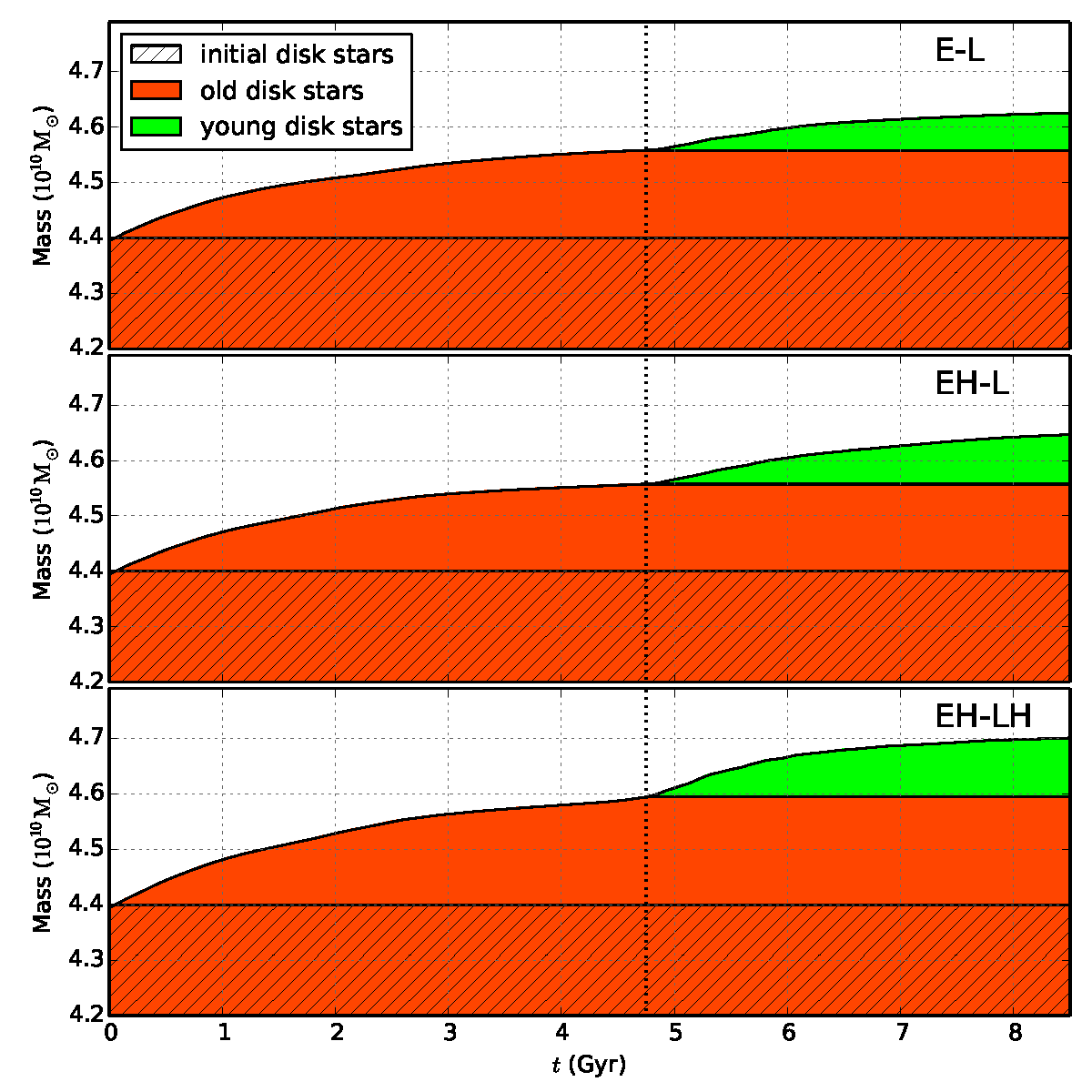}
\caption{
The growth of the total mass 
of the stars on the disk of the LTG together with those transferred 
to the ETG (if there are any) 
in the three runs.
In each panel, the orange shade represents 
the total mass of the old disk stars, 
i.e., initial disk stars plus stars formed before 
the closest approach. 
The diagonal lines are drawn to show the total mass 
of the initial disk stars, which is 
4.4~$\times$ $10^{10}\,\rm{M_{\odot}}$ in all runs. 
The closest approach time is marked with 
the vertical dotted line at $t =$ 4.75~Gyr. 
The total mass of the young stars that have been formed 
since the closest approach are shown with the green shade. 
}
\end{figure*}

\begin{figure*}[!hbt]
\centering%
\includegraphics[width=14.5cm]{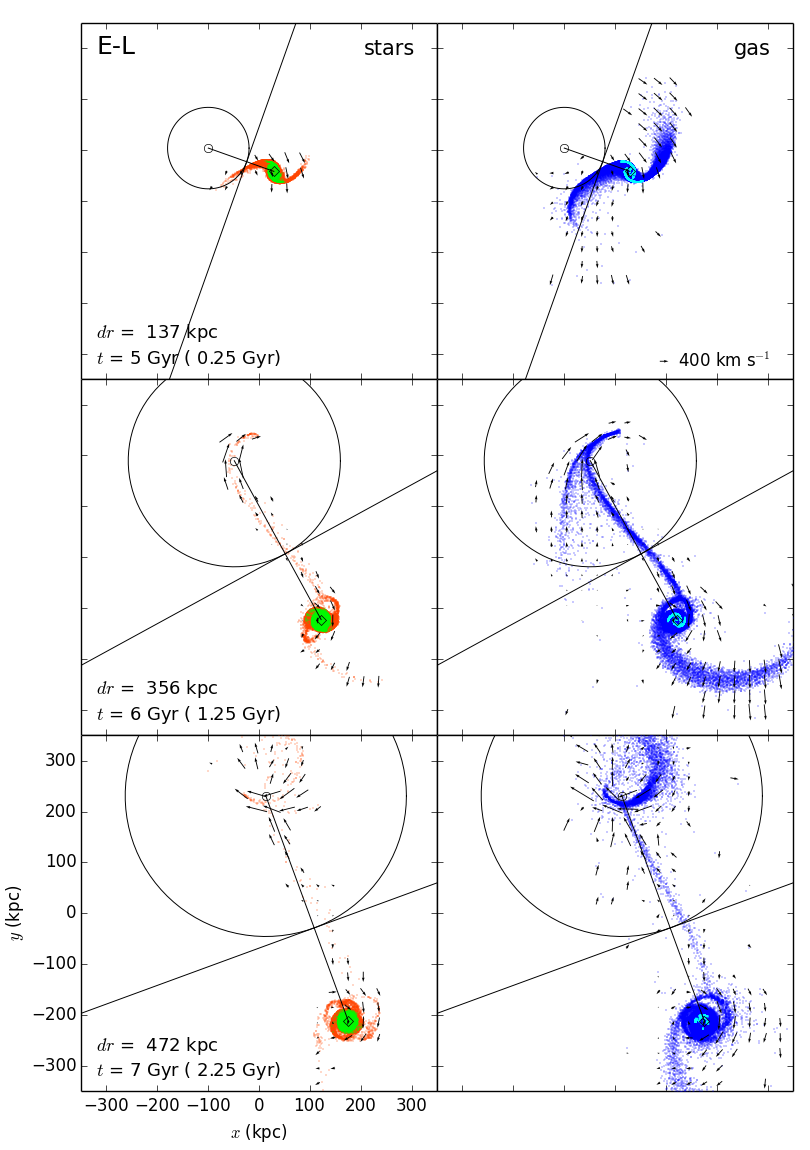}
\caption{ 
Disk materials in run~E-L with the velocity vectors (relative to the early-type galaxy) 
projected onto the $x$$-$$y$ plane 
at $t = $ 5, 6, and 7~Gyr (from the top to the bottom rows, respectively).
The disk stars and gas particles are shown in the left and the right columns, respectively.
The separations between the two galaxies 
and the times are shown in the left panels. 
The times written in the parentheses are those measured 
from the closest approach. 
In each panel, the colors are used in the same way as in Figure~2 
(orange for the old disk stars, 
green for the young disk stars, blue for the non-star-forming disk gas, 
and cyan for the star-forming gas).
The arrows represent the transverse velocities of the disk particles 
in a reference frame centered on the center of the ETG.
The lengths of the arrows are proportional to the 
magnitude of the transverse velocities.
The center-of-mass positions of the early-type and 
the late-type galaxies are marked 
with a small circle and a small diamond, respectively.
A line connecting between the centers of the two galaxies is drawn.  
On the line, at the position where the net local gravitational acceleration 
becomes zero, assuming each galaxy as a point mass, 
another line is drawn perpendicular to the line connecting the centers. 
At later times (after $t~\sim$~5.5~Gyr), 
the particles lying within the large spheres 
inscribed in the perpendicular line 
are considered 
to be captured by the early-type galaxy.
}
\end{figure*}

\begin{figure*}[!hbt]
\centering
\includegraphics[width=16cm]{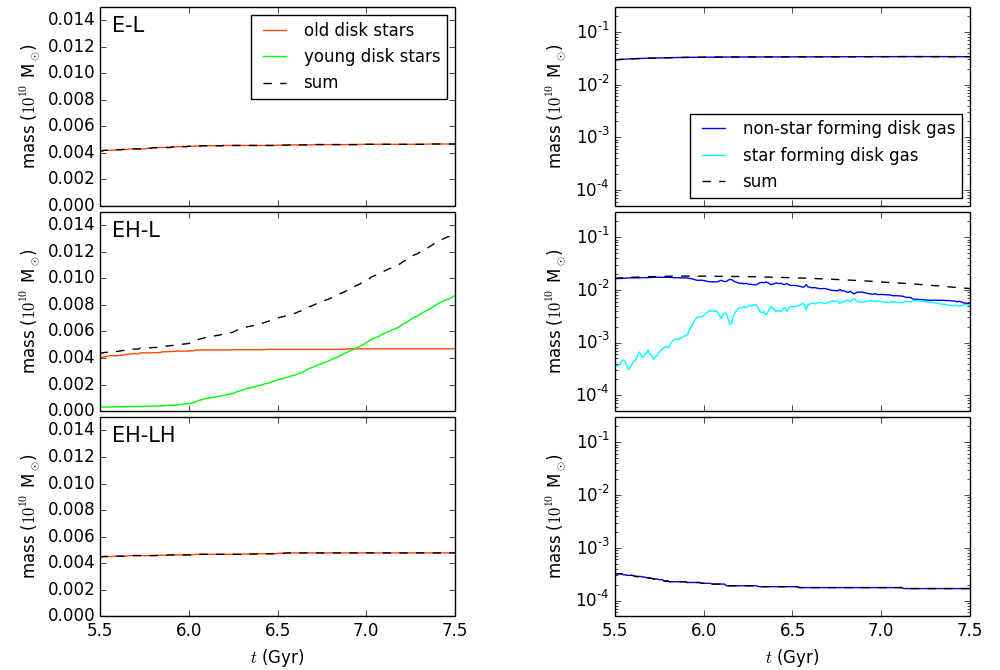}
\caption{Disk materials transferred to the early-type galaxy 
within the spheres shown in Figures~4, 6, and 8 in runs~E-L, 
EH-L, and EH-LH (from the top to the bottom rows, respectively).
The left panels show the amount of the disk star particles 
in units of $10^{10}\,\rm{M_{\odot}}$. 
The amounts of the old disk stars 
and the young stars formed since the closest approach 
are displayed separately, 
with the orange and the green solid lines, respectively.
The total amount of the old and the young stars is 
shown with the black dashed line.
The right panels present the amounts of the 
non-star-forming disk gas (blue solid line), 
star-forming disk gas (cyan solid line), 
and the sum of the two (black dashed line) 
in log scale.   
In both runs~E-L and EH-LH, 
no young stars or star-forming gas 
transfer to the early-type neighbor. 
}
\end{figure*}


\section{Results of distant encounters}


\subsection{Orbital Parameters and Trajectories}

We run three simulations of distant encounters between 
the ETG and the LTG 
to examine the effects of the hot gas halo.
They are the E-L, EH-L, and EH-LH runs. 
Their initial conditions are listed in Table~2. 
In all runs, the mass ratio of the ETG to LTG is 2:1 (Table~1). 
(We have also considered the E-LH case. 
This run is presented in Appendix~B.)

In each run, the ETG is initially placed at the origin 
and the LTG is positioned 
at ($x_{0},y_{0},z_{0})=(-1200~\rm{kpc}, 265~\rm{kpc}, 0)$ 
with the initial relative velocity 
of ($v_{x0},v_{y0},v_{z0})=(200~\rm{km}~s^{-1}, 0, 0)$ 
toward the ETG. 
The initial $y$ position (initial impact parameter) 
of the LTG is chosen 
so that 
the closest approach distance 
becomes about 100~kpc. 
(Throughout this paper, we consider  
the position of the center of mass of the bulge of each galaxy 
as the center position of the total galactic system.)
The disks of all LTGs are initially set  
in the $x$$-$$y$ plane 
without tilt and have clockwise directional spin 
when viewed from the 
positive $z$~axis.

The three simulations  
are run for over 8~Gyr.
In the left panel of Figure~1, 
the distance 
between the LTG and the ETG, $dr$, 
and the relative velocity of the LTG, $dv$, 
in run~E-L 
are plotted as a function of time.
The two galaxies encounter most closely at 
$t = 4.75$~Gyr with $dr = 94$~kpc.
(In this paper, time $t$ represents the time elapsed 
since the start of the simulation, if not otherwise specified. 
The projected distance between the galaxies onto the $x$$-$$y$ plane 
is close to the three-dimensional distance $dr$ throughout the run 
because the galaxies move mostly in the $x$$-$$y$ plane.)
In the right panel of Figure~1, 
the orbital trajectories 
of the late-type (thick solid line) 
and the early-type (thin solid line) galaxies in run~E-L are presented.
The positions of the two galaxies 
at $t = $ 4, 5, 6, and 7~Gyr are marked on their trajectories 
with circles, crosses, triangles, and squares, respectively.

As for runs~EH-L and EH-LH, 
the orbital trajectories of the galaxies 
are almost the same as those in run~E-L 
(see Table~2 for the closest approach events).


\subsection{Evolution of Run~E-L}

Figure~2 (top two rows) shows the snapshots 
taken at $t = $~4, 5, 6, and 7~Gyr.
The closest approach is at $t = 4.75$~Gyr. 
The time measured from the closest approach (written in the parentheses) 
and the three-dimensional separation between the two galaxies, $dr$, 
at each time are given in the lower panels.

The second snapshot shows early development of 
a tidal bridge and a tail out of 
the stellar and the gas disks of L shortly after the closest approach. 
The third snapshot shows that 
non-star-forming disk gas (blue points) and 
``old" disk stars 
(orange points; the disk stars existed before the closest approach) 
are transferred from L to E. 
The cold non-star-forming gas flows through the tidal bridge 
and dynamically follows the motion of the accreted old stars. 
However, none of the green points, which represent the stars formed 
out of the gas since the closest approach 
(hereafter, ``young" (disk) stars), nor any of the cyan points, 
which represent the gas particles having positive values 
of SFR at that instant, i.e. star-forming gas,  
are transferred to E in this run. 
In the last snapshot, 
the accreted cold gas and 
the old stars appear 
to orbit around the center of E. 
(Throughout this paper, we refer to the particles by their original setup, 
if not otherwise specified. 
For example, ``disk star particles of the LTG" mean the particles 
initially set as the disk star particles of the LTG. 
Similarly, ``disk gas particles" mean the particles initially set as the 
disk gas and those have not turned into stars by the time.) 
The total mass of the old and young disk stars 
is shown in Figure~3 (top panel).

To show the mass transfer from L to E more clearly, 
we display in Figure~4 the disk particles of L together 
with the velocity vectors 
in a reference frame centered on the center of the ETG 
projected onto the $x$$-$$y$ plane. 
The disk star and gas particles move toward E through the bridges 
in a dynamically similar fashion. 
To estimate the amount of the disk particles transferring to E, 
we use a sphere centered on E with the radius to the point where the 
net local gravitational acceleration becomes zero along the line connecting 
the centers of E and L, 
assuming each galaxy as a point mass. 
At $t = $~6~Gyr, some of the old disk stars (orange points) 
and a relatively large number of 
cold non-star-forming disk gas (blue points) 
are transferring to E, seen within the sphere; 
however, none of the young stars (green points) 
nor any of the star-forming disk gas (cyan points) 
are found within the sphere. 
By $t = 7$~Gyr, old disk stars and cold disk gas 
have continued to transfer to E.
Those materials captured by E move around the center of E.
The star and gas bridges 
begin to fade away. 
Figure~5 (top panels) presents the total mass of the disk materials 
enclosed within the spheres with respect to time 
after the closest approach 
(for $t \geq 5.5$~Gyr).    
At $t$~=~6~Gyr, 
about $0.45~\times$ $10^{8}\,\rm{M_{\odot}}$ of old disk stars 
and $3.3~\times$ $10^{8}\,\rm{M_{\odot}}$ of non-star-forming disk gas 
are accreted onto E. 
The amount of the transferred cold gas is about 
5\% of the original cold disk gas mass.
The accreted materials do not include young disk stars 
or star-forming gas in this run.


\subsection{Evolution of Run~EH-L}

As shown in the snapshots in Figure~2 (third and fourth rows), 
the distribution of the old disk stars (orange points in the upper panels) 
is generally the same as 
that in the E-L run throughout the simulation. 
However, the young disk stars (green points) 
and the disk gas (blue points) 
show very different configurations. 
In the first snapshot at $t = 4$~Gyr,  
the gas disk forms a bow-like front as it moves against 
the halo gas included in EH (red points in the lower panels), 
and some disk gas at the other side is stripped. 
In the next snapshot, at 0.25~Gyr after the closest approach, 
the gas bridge develops nearly straightly, 
while the stellar bridge consisting of the old disk stars 
bends below the gas bridge. 
There is a burst of star formation (green points) 
in the shock-compressed gas along the gas bridge 
right after the closest approach. 
The offset between those young stars and the old stars 
reflects the strength of the shock.
The gas bridge, including some of star-forming gas (cyan points), 
continues to extend almost radially to the center of EH. 
The last snapshot shows that 
the accreted old and young stars 
orbit around EH with different orbits. 
Some young stars and cold gas accumulate at the center of EH. 
It should be noted that, in the previous run E-L, 
no young stars manage to transfer to 
the ETG (see the top two rows in Figure~2).
The middle panel of Figure~3 shows overall more production 
of young stars (including those transferred to EH) 
in the EH-L~run than in the E-L~run.

The disk particles are displayed separately in Figure~6. 
The total masses of the disk gas and disk star particles 
transferred to EH are shown in Figure~5 (middle row).
Compared to run~E-L, 
a similar amount of the old disk stars but 
less than half the amount of non-star-forming cold gas 
is transferred to EH within the large spheres drawn in Figure~6.
At $t$ = 5~Gyr (top row in Figure~6), 
the young stars (green points) and the star-forming gas (cyan points) are seen 
over a wider area along the bridges compared to those in run~E-L; 
some of them are located beyond the line of the zero gravity toward EH. 
At later times, EH is able to accrete 
the star-forming gas and some of the young stars to its center. 
In addition, some non-star-forming gas particles 
accumulated at the center of EH 
later turn into stars in this run.

In Figure~7, 
we present the local density $\rho_{\rm gas}$ and 
the temperature $T$ of the gas particles 
located near the orbital plane (within ${\mid}z{\mid} \le 15$~kpc) 
to show how cold disk gas is affected 
by hot halo gas through the hydrodynamic interaction. 
As the gas disk approaches the gas halo (top row), 
the gas density at the leading side of the disk rises, 
and some halo gas piles up along the bow-like front of the disk.
The piled gas has higher temperatures due to the shock heating.
Shortly after the closest approach (middle row), 
a channel for the cold gas toward the center of 
E is created. 
This channel is seen between the inner trailing shock 
wound around EH taken over by the supersonic motion of L 
and the preceding shock that appears due to the cold gas tidal bridge. 
The cold gas can flow though this channel 
and is not blocked by shocks because the shocks exist parallel to
the cold gas bridge. 
(We have checked that the positions of the shocks 
in this run and in run EH-LH, 
which strongly determine the subsequent evolution, 
are unchanged in different-resolution runs.
We present the high-resolution simulations in Appendix~C.)

\begin{figure*}[!hbt]
\centering
\includegraphics[width=14.5cm]{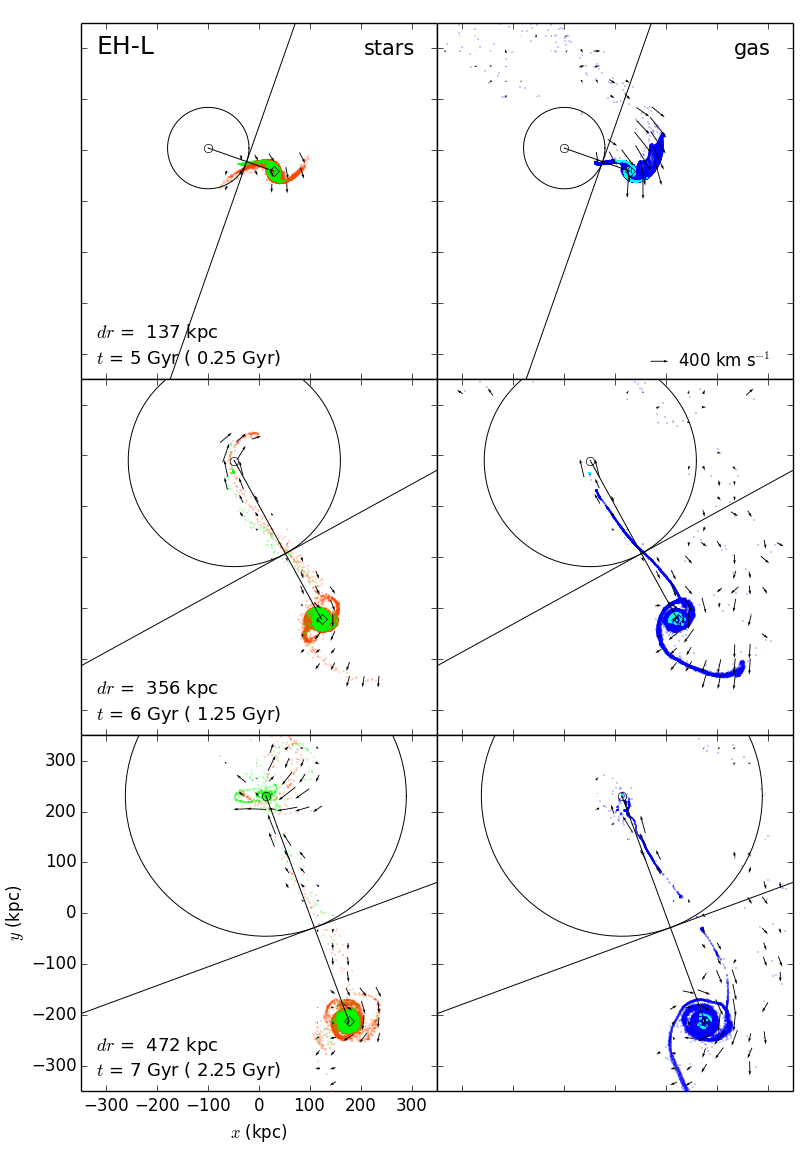}
\caption{\centering The same as Figure~4, but for run~EH-L. 
}
\end{figure*}

\begin{figure*}[!hbt] 
\centering
\includegraphics[width=15.5cm]{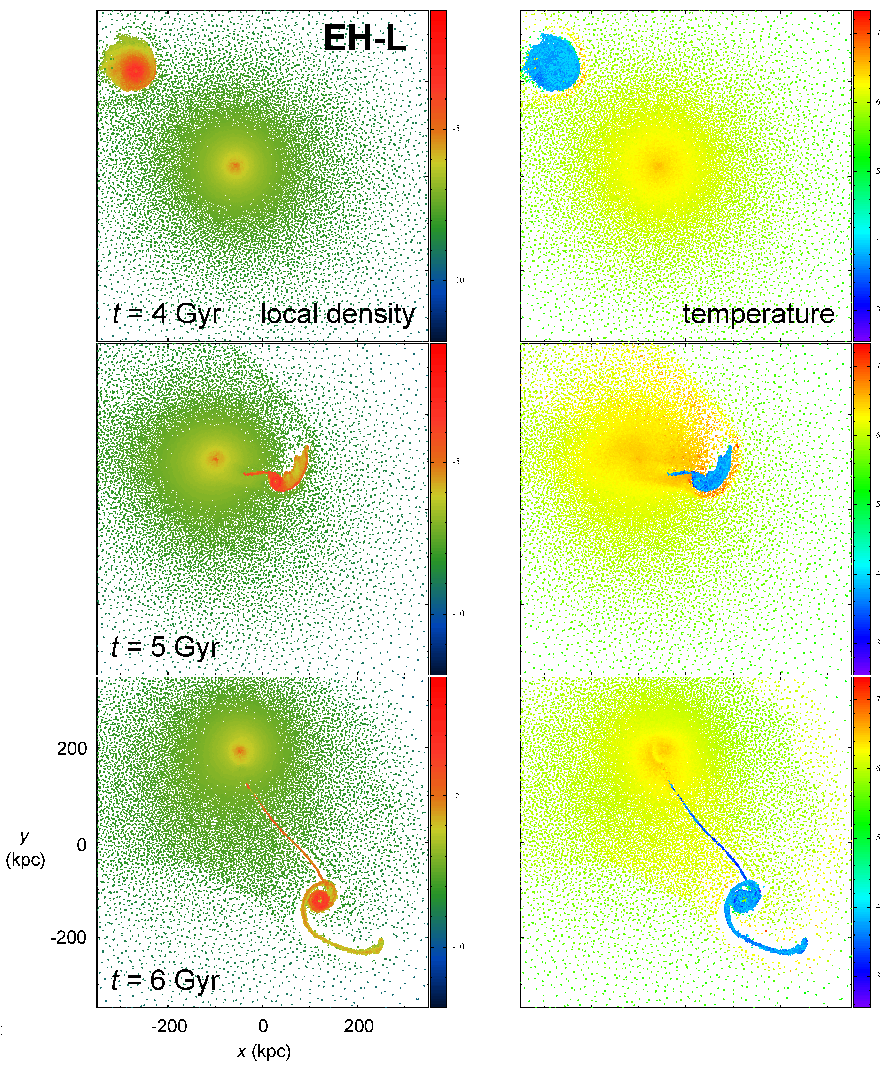}
\caption{The local density and the temperature 
of the gas particles in run~EH-L 
at $t =$ 4, 5, and 6~Gyr. 
In each panel, the gas particles of both early- and late-type galaxies near the 
orbital plane, with ${\mid}z{\mid} \le 15$~kpc, are displayed.  
The colors of the gas particles represent the values 
of $log\,\rho_{\rm gas}$ in $10^{10}\,\rm{M_{\odot}}$~kpc$^{-3}$ (left column) 
and $log\, T$ in K (right column). 
The ranges of the color bars in each column are fixed.
This figure is made by using SPLASH (\citealt{Price2007}). 
}
\end{figure*}

\begin{figure*}[!hbt]
\centering
\includegraphics[width=14.5cm]{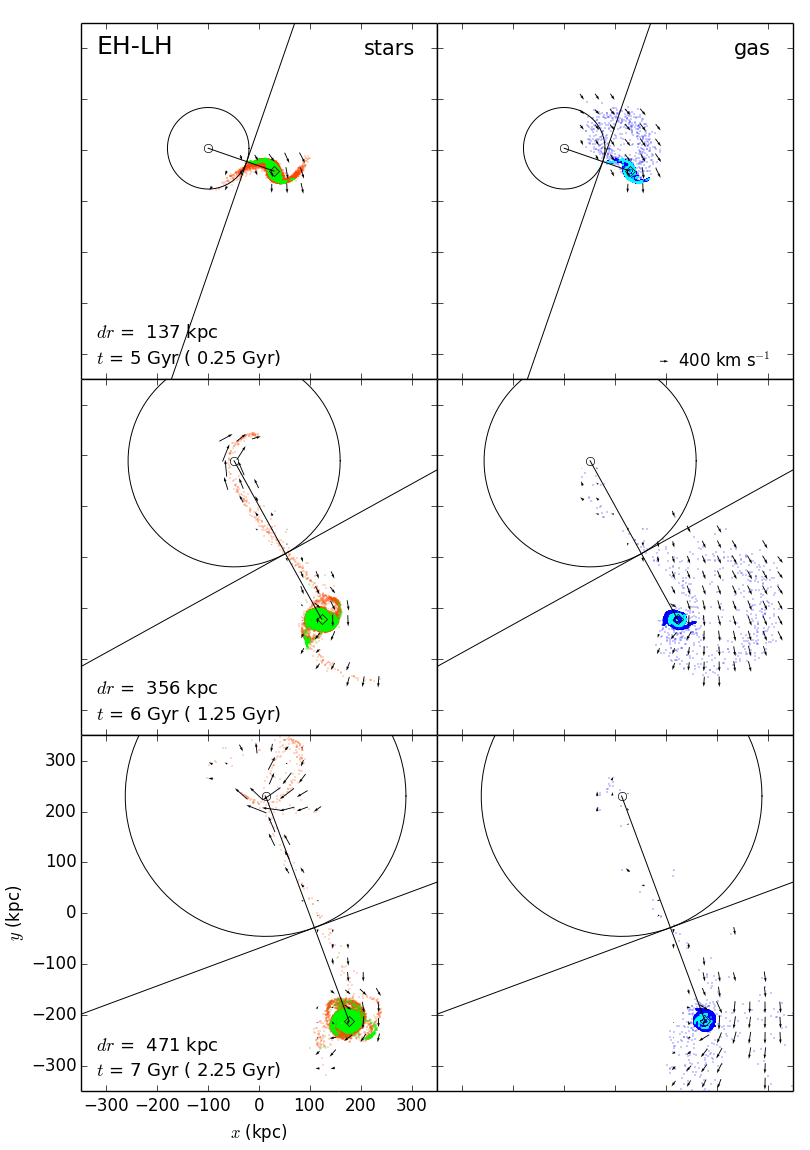}
\caption{\centering The same as Figures~4 and 6, but for run~EH-LH.
}
\end{figure*}

\begin{figure*}[!hbt] 
\centering
\includegraphics[width=15.5cm]{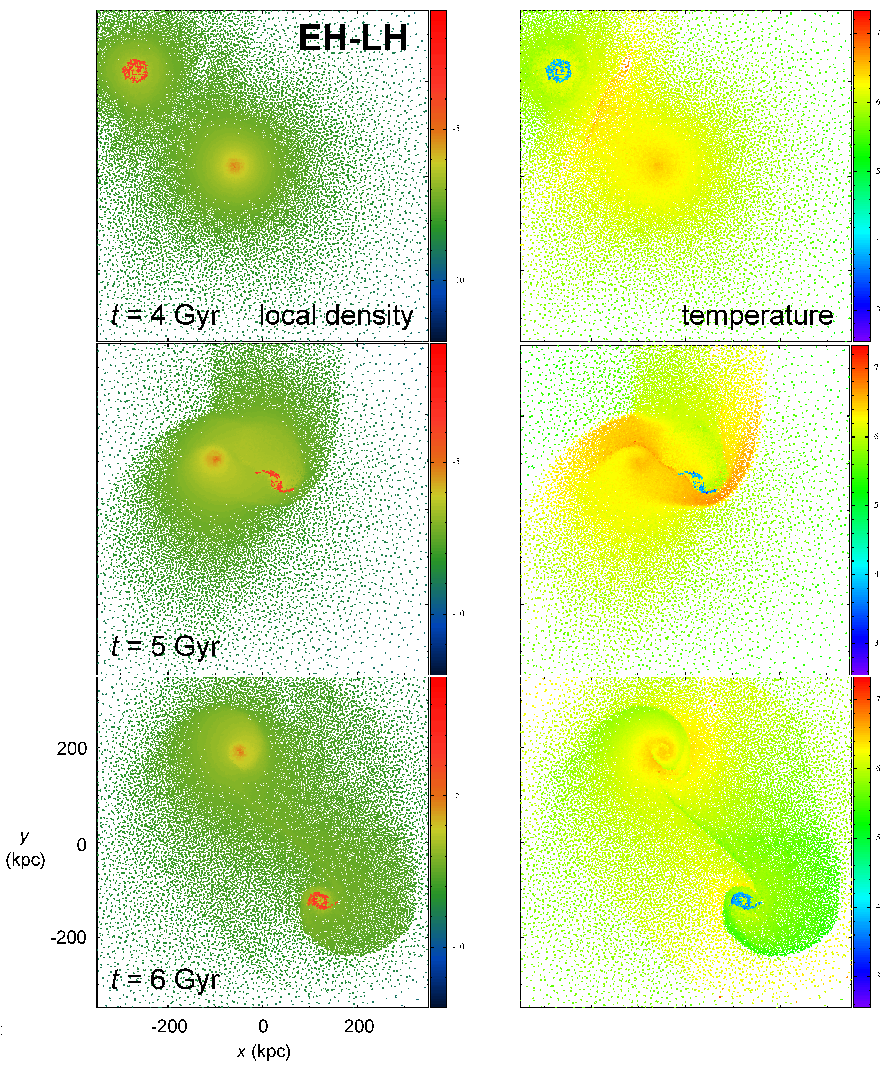}
\caption{\centering The same as Figure~7, but for run~EH-LH. 
}
\end{figure*}

\begin{figure*}[!hbt]
\centering
\includegraphics[width=14.5cm]{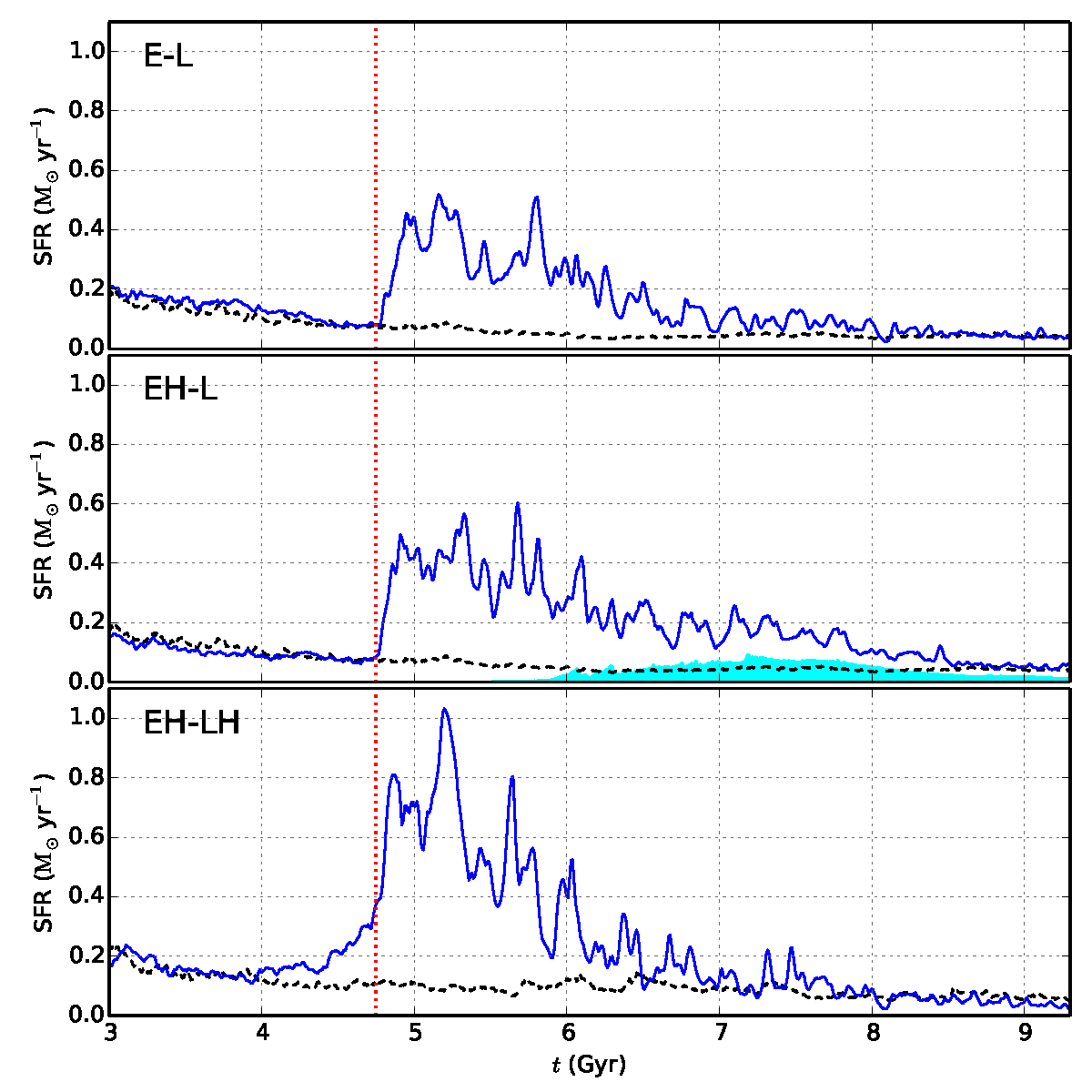}
\caption{Evolution of star formation rate 
in runs~E-L, EH-L, and EH-LH 
(from the top to the bottom panels, respectively). 
In each panel, the solid blue line represents the sum of the SFRs 
of all gas particles that initially belong to the late-type galaxy; 
the dashed black line is the sum of the SFRs of all gas particles 
of the late-type galaxy in isolation.
The time of the closest approach 
is marked with the vertical dotted line (red).  
In the middle panel, 
the shaded region with cyan color  
denotes the contribution to the total SFR 
(shown with the blue solid line) 
from the gas particles transferred to the early-type galaxy 
(within the spheres shown in Figure~6) 
since $t =$~5.5~Gyr. 
There is no cyan shaded region in the top and the bottom panels 
because none of the star-forming 
gas particles are captured by the early-type neighbor. 
In the bottom panel, the SFRs of the gas particles originally 
set as the halo gas of LH are included in both the blue and black lines, 
but the contribution is very small (see the text).
}
\end{figure*}

\begin{figure*} 
\centering
\includegraphics[width=16cm]{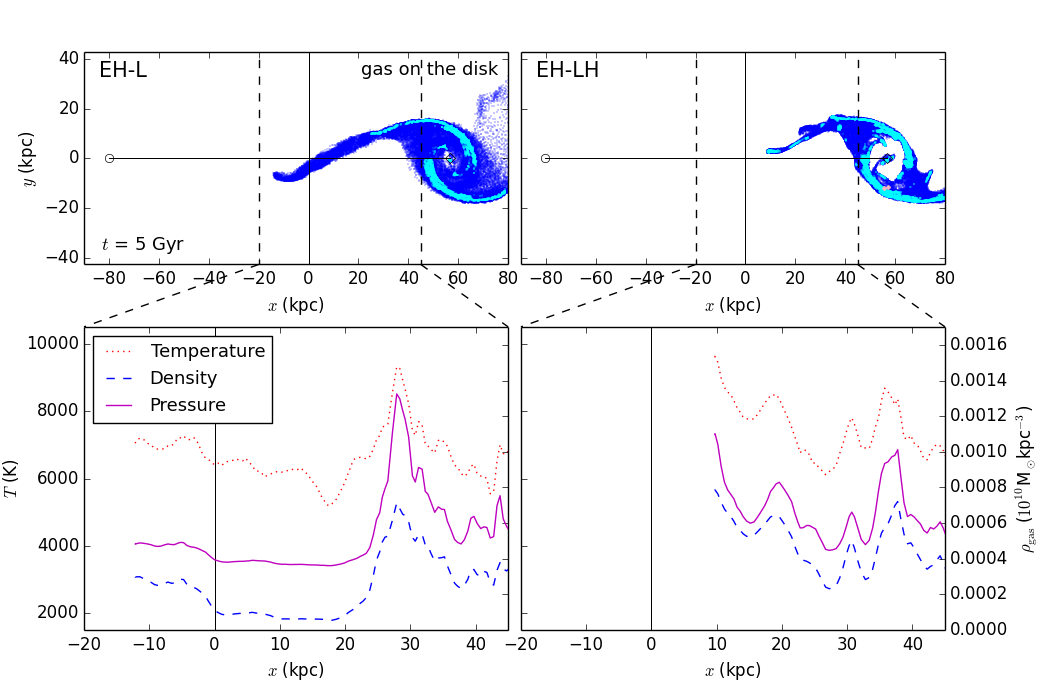}
\caption{
{\it{Top row}}: 
gas particles on the disk at $t$ = 5~Gyr 
from runs~EH-L (left) and EH-LH (right).
The particles originally set as the disk gas 
but stripped off or scattered into the halo 
are not shown. 
The blue and cyan points represent non-star-forming 
and star-forming gas particles, respectively. 
The pink points seen in the right panel are 
the original halo gas particles of LH 
that cooled down onto the disk having a zero star formation rate. 
The positions of the center of mass of the early- and 
the late-type galaxies 
are marked with a circle and a diamond, respectively. 
The horizontal axis is set along the line connecting 
the centers of the two galaxies.  
The zero point on the $x$ axis is chosen to be 
the position where the net local gravitational acceleration 
due to the two galaxies becomes zero. 
{\it{Bottom row}}: 
average profiles of the temperature $T$ (red dotted line), 
local density $\rho_{gas}$ (blue dashed line), 
and pressure $P$ (magenta solid line) 
for the gas particles around the bridge area, 
located between the two vertical dashed lines drawn in the top panels. 
To obtain the average quantities, 
the gas particles shown in the top panels are projected 
onto the $x$ axis, and then the average 
is calculated in every 0.5~kpc along the $x$ axis, 
taking $N$ nearest particles into account to the center of each bin. 
The number of particles $N$ is chosen 
to be five times the average number of gas particles in the bridge 
within a 0.5~kpc length on the $x$ axis. 
Each of the three gas quantities in the left and the right panels 
is displayed on the same scale: 
the temperature scale is shown to the left of the left panel 
and the local density scale is shown to the right of the right panel. 
The pressure scale is chosen to be fit in the same panel.   
The gas bridge in run~EH-L (left) is more extended than that in run~EH-LH (right) 
beyond the vertical solid line at $x$ = 0.
}
\end{figure*}


\subsection{Evolution of Run~EH-LH}

Figure~2 (bottom two rows) shows the four snapshots 
from run~EH-LH, 
where both ETG and LTG 
possess hot gas in their surrounding halos 
(red and pink points, respectively). 
The old disk stars evolve in the same way as in the 
two previous runs at all times, 
but the cold gas evolves in a unique way, being affected by the 
gas halos. 
At $t = 4$~Gyr, a large-scale shock is formed 
between the two hot gas halos (bottom-left panel).
By this time, some disk gas 
at the outskirt has been ionized and scattered into 
the halo because of the hot halo gas of its own galaxy. 
In addition, the gas disk in this run 
has dissipated more quickly than the other two runs 
and produced more stars 
because of the compression exerted by its halo 
(see Paper~I for more details), 
before the gas halo of LH is 
affected by the collision 
with the gas halo of EH.
At later times ($t$~=~5 through 7~Gyr), 
the gas disk does not develop a strong bridge as in the other runs, 
and almost no cold gas is transferred 
to EH.\footnote{The strength 
of a gas bridge will vary in different settings. 
For example, in the case of closer encounters, 
a gas bridge might develop more strongly 
to the neighboring galaxy.}
Some halo gas particles of LH cool down to the disk and 
then turn into stars. 
The total mass of these stars formed 
out of the original halo gas of LH 
is less than $\sim$~1~\% of that of the stars generated 
out of the original disk gas; 
they are found at the central part of the disk.
Figure~3 (bottom panel) shows that 
the total mass of the young stars increases rapidly 
near the closest approach 
but slows down after a couple of gigayears. 
These young stars are all found on the disk of LH, 
not within EH.

Figure~8 displays the distribution of the disk particles.
At $t =$ 5~Gyr, the young stars formed since the closest approach 
(green points) appear along the bridge, 
but are less extended compared to run~EH-L; 
the gas bridge does not grow
beyond the line of zero gravitational acceleration.
At $t =$ 6 and 7~Gyr,
some old stars (orange points) are seen transferred to EH. 
However, no young stars and 
only a little non-star-forming gas (blue points) are captured by EH. 
The total masses of the star and gas particles 
transferred to EH 
are presented in Figure~5 (bottom row).
The amount of the old stars transferred to the early-type neighbor 
is similar to (a little bit more than) those in the previous runs. 
But the total mass of the cold gas moved to EH 
is much less than that in the other two runs.

The local gas density and the gas temperature 
are shown in Figure~9.
The discontinuities of the gas quantities 
are seen at the shock. 
As LH turns around EH (middle row), 
the inner trailing shock propagates toward the center of EH 
before the cold gas tidal bridge starts to extend toward EH, 
blocking its motion to the center of EH. 
While the cold disk has been already dissipated a lot by this time  
through star formation, 
the blocking by the shock further hinders 
the growth of the cold gas bridge and ionizes it. 
The shock is later completely wound around EH 
and makes a spiral pattern in the hot halo gas (bottom row).


\subsection{Star Formation Activities}

We present the SFRs in the three runs 
in Figure~10.
In run~E-L (top panel), 
the SFR (blue solid line), 
which is the sum of the SFRs from all gas particles 
originally set as the disk gas, 
rapidly increases 
right after the closest approach 
at $t = 4.75$~Gyr (marked by a vertical dotted line). 
This compares with a gradual decline of SFR of 
the corresponding isolated L (black dashed line). 
The value at the first peak of the blue line at $t$~=~4.95~Gyr 
is 0.46 $\rm{M_{\odot}}~yr^{-1}$, 
which is about six times the value of the isolated galaxy 
at the same instant. 
The excess of the SFR 
lasts until $t\sim$~8~Gyr.

In run~EH-L (middle panel), 
the SFR (blue line) also abruptly exceeds 
that of the isolated galaxy~L (black line) 
right after the closest approach.
The excess is a little greater and lasts longer than that in run~E-L. 
The SFR at the first peak is 0.5~$\rm{M_{\odot}}~yr^{-1}$ 
at $t = 4.91$~Gyr, which is about seven times that 
of the isolated galaxy~L. 
(We neglect the SFRs of the halo gas 
particles of the ETG model EH 
because they are very small, although not always zero, 
with a maximum value of 0.018 $\rm{M_{\odot}}~yr^{-1}$).
This implies that the enhanced star formation in this run 
is due to the compression of the gas disk  
through the collision with the gas halo of EH.
Although some of the disk gas particles 
are heated and stripped, 
the compression by the shock initiated during the collision
is mainly responsible for the cold gas having a higher SFR. 
Among the star-forming gas particles of L, 
some are captured by EH 
after the encounter.
The total SFR of the gas particles transferred to EH 
is shown with the cyan shade for $t \geq 5.5$. 
Some non-star-forming disk gas transferred to EH 
also turns into star-forming gas particles later, 
when they gather densely enough to satisfy 
the conditions for star formation. 
These particles lead to the increase of 
the shaded region around $t$ = 7~Gyr.

In run~EH-LH (bottom panel), 
the SFR (blue line) also exceeds 
that of the isolated galaxy~LH (black line) 
but much more greatly than in the other two runs, 
with the SFR of 0.81~$\rm{M_{\odot}}~yr^{-1}$ 
at the first peak at $t = 4.87$~Gyr.
The maximum value of the SFR reaches to 
1.033~$\rm{M_{\odot}}~yr^{-1}$ at $t = 5.2$~Gyr. 
Unlike in the others, 
the excess (over the black line) 
begins earlier at about $t$ = 4~Gyr, somewhat before 
the closest approach, 
from the onset of the collision between the two gas halos.
The shock-compressed gas halo of the LTG 
consequently delivers the effects onto the cold disk, 
raising the SFR since $t \sim$~ 4~Gyr. 
The great excess on the SFR near the closest approach 
indicates the strong compression from the shock generated by the collision. 
Compared with the SFR of the isolated L, 
the SFR of the isolated LH is generally higher, 
mainly due to the compression exerted by the gas halo of LH; 
the direct contribution from the original halo gas particles of LH 
to the total SFR 
is small (at most about $1\%$ or less in the total SFR of LH). 
\\

\section{SUMMARY AND DISCUSSION}

Using $N$-body/SPH simulations, 
we studied distant encounters 
between an ETG and an LTG 
for a case of the closest approach distance of about 100~kpc 
with the mass ratio of the ETG to LTG of 2:1. 
We performed three comparison simulations that adopt galaxy models 
having or not having a hot gas halo, 
keeping all of the other parameters fixed.
We summarize our key findings in each run as follows, 
focusing on the mass transfer of the disk materials to the ETG 
and the star formation enhancement that is due to the encounter.

1. E-L Case\\
A large amount of cold gas is transferred from L to E, 
but it does not contain star-forming gas.  
The cold gas flows through the gas bridge and 
dynamically follows the motion of accreted old disk stars. 
Because there is no compression of the cold gas by shock, 
no new stars form along the tidal bridge.
Thus, the ETG accretes cold gas and old disk stars 
but no young stars nor star-forming gas out of the LTG. 
The accreted cold gas and old disk stars 
orbit around the center of E. 
The SFR of L in run~E-L increases abruptly 
right after the closest approach.

2. EH-L Case\\
Near the closest approach, 
a shock forms along the cold disk gas tidal bridge, and 
new stars form along the bridge. 
The strength of this shock is seen from the offset 
between the tidal bridge that consists of the old disk stars 
and that consisting of the young stars and the cold gas. 
A very important feature of this type of interaction is that 
a channel for the cold gas toward the center of EH is created 
right after the closet approach.  
This channel is developed between 
the inner trailing shock wound around EH 
taken over by the supersonic motion of L 
and the preceding shock that appears due to the cold gas tidal bridge. 
The cold gas can flow almost radially through this channel 
and is not blocked by shocks 
because the shocks exist parallel to the cold gas tidal bridge. 
The cold gas flowing to EH contains some star-forming gas. 
Some of the non-star-forming gas transferred to EH 
later turns into stars as well.
As a result, 
EH can accrete the cold disk gas, including some star-forming gas, 
\emph{at its center} in this particular run, 
and also accrete the old and young disk stars 
\emph{around the center}. 
Some of the young stars are found at the center, as well.
The SFR of L in this run also increases rapidly 
right after the closest approach.
The increase is greater than that in run~E-L. 
Then, the SFR decreases most slowly among all three runs 
mainly because the accreted cold gas turns into stars later 
at the center of EH.

3. EH-LH Case\\
As LH approaches EH, a shock is formed between the two hot gas halos. 
The cold disk gas is compressed by the hot halo and produces more stars than 
the other two runs before the closest approach. 
The dissipation of the cold gas occurs much more effectively 
than the accretion of the halo gas onto the disk.  
When LH swings around EH near the closest approach, 
the inner trailing shock propagates toward the center of EH 
before the tidal bridge develops 
out of the gas disk, which has already been largely dissipated, 
toward EH and blocks its motion to the center of EH and ionizes it. 
Therefore, only the old disk stars flow along the tidal bridge, 
and there is no burst of star formation within the cold
gas that can be accreted to EH. 
The result of the interaction is mostly the old disk star 
accretion from LH to EH. 
The SFR of this run arises somewhat before the closest approach 
because of the shock between the two gas halos.
The maximum value of the SFR after the closest approach 
is much greater than that in the other two runs, 
but it decreases rapidly because of the lack of the cold gas.

The almost no gas accretion from the LTG to the ETG in run~EH-LH, 
unlike in run~EH-L,  
is hinted at in the snapshots 
taken shortly after the closest approach 
when the gas bridge develops. 
We show in Figure~11 the gas bridges 
in runs~EH-L and EH-LH at $t$ = 5~Gyr 
more closely. 
In the top panels, 
the gas particles within the disks are displayed\footnote{
The original disk gas particles that are stripped off 
or scattered into the halo are not plotted. 
Those particles that joined the halo 
are easily distinguished on the $\rho_{gas}$$-$$T$ plane, 
as explained in Paper~I.} on a new $x$$-$$y$ plane 
that has an $x$~axis 
along the line connecting the centers of the two galaxies 
and a zero point at the position 
where the net local gravitational acceleration 
is zero. 
The gas bridge in run~EH-L (left panel) develops toward the ETG, 
well beyond the $x$ = 0 line at the time. 
However, in run~EH-LH (right panel), the gas bridge 
does not grow well, reaching behind the $x$ = 0 line. 
In order to transfer some gas to the ETG in run~EH-LH, 
the gas bridge should have grown beyond the zero-gravity line 
before it is blocked by the shock. 
In addition,  
the pressure of the gas particles, 
shown in the bottom-right panel (magenta solid line), 
rises near the end of the bridge.  
This implies that those particles 
feel the net force (gravity plus pressure gradient force) 
toward the LTG and  
will not be able to flow to the ETG at later times.

The above results 
would vary in different situations. 
For example, in a case where LH encountered with EH more closely 
(keeping the other orbital parameters unchanged), 
the gas tidal bridge might be able to grow more strongly to the ETG 
before being blocked by the shock. 
Then, some gas will manage to flow to the ETG despite the shock.  
The materials transferring through the gas 
and stellar bridges 
might include star-forming gas and young stars in run~EH-LH, 
as in run~EH-L.
It is also possible for the non-star-forming cold gas to flow 
to the ETG for a while 
through the gas bridge and turn into stars there 
when they become dense enough.     
The gas bridge is expected to vanish more quickly 
in run EH-LH after all, compared with the gas bridge in
run EH-L, because 
the shock would still cross the gas bridge and ionize it. 
In the case of EH-L, 
the SFR might start to increase before the closest approach, 
as in run~EH-LH, 
when they encountered each other more closely.     
Finally, in run E-L 
we expect that the disk materials captured by the ETG 
will hardly include star-forming gas and young stars, 
even in somewhat closer encounters 
because there is still no compression of the cold gas by shock.
The materials transferring to E through the star and gas bridges 
would also more likely orbit around the center of E 
and rarely settle down at the center, 
unless all of the model and orbital parameters were matched 
to cancel out the transverse component of the velocities 
of the particles, 
leaving only the radial component. 
Some cold gas transferred to E might be able to turn into stars 
later when it becomes dense enough in closer encounters.

Overall, our simulations show that 
the cold gas flow and 
the star formation activity 
depend strongly on 
whether or not the hot gas halo is included 
in the ETG and/or the LTG, 
whereas the amount of the old disk stars accreted to the ETG 
and their dynamics are almost the same in all runs.  
Therefore, we claim that it is necessary to include 
hot halo gas properly 
in galaxy interaction simulations in order to 
more accurately study the transient phenomena.


\begin{acknowledgments}
We thank the referee very much for providing insightful comments 
that improved this paper. 
J.-S.~H is grateful to C. Struck and J.-H. Choi 
for helpful discussions 
and to B. Cervantes-Sodi, B. L'Huillier, 
and X. Li for help in making some of the figures. 
We appreciate 
Joshua E. Barnes making the ZENO code available and 
Volker Springel for providing us with GADGET-3. 
We thank the Korea Institute for Advanced Study
for providing computing resources (KIAS Center 
for Advanced Computation 
Linux Cluster System) for this study. 
J.-S.~H acknowledges the support by the BK21 plus program 
through the National Research Foundation (NRF) funded by 
the Ministry of Education of Korea.
\end{acknowledgments}

\begin{figure*} 
\centering
\includegraphics[width=16cm]{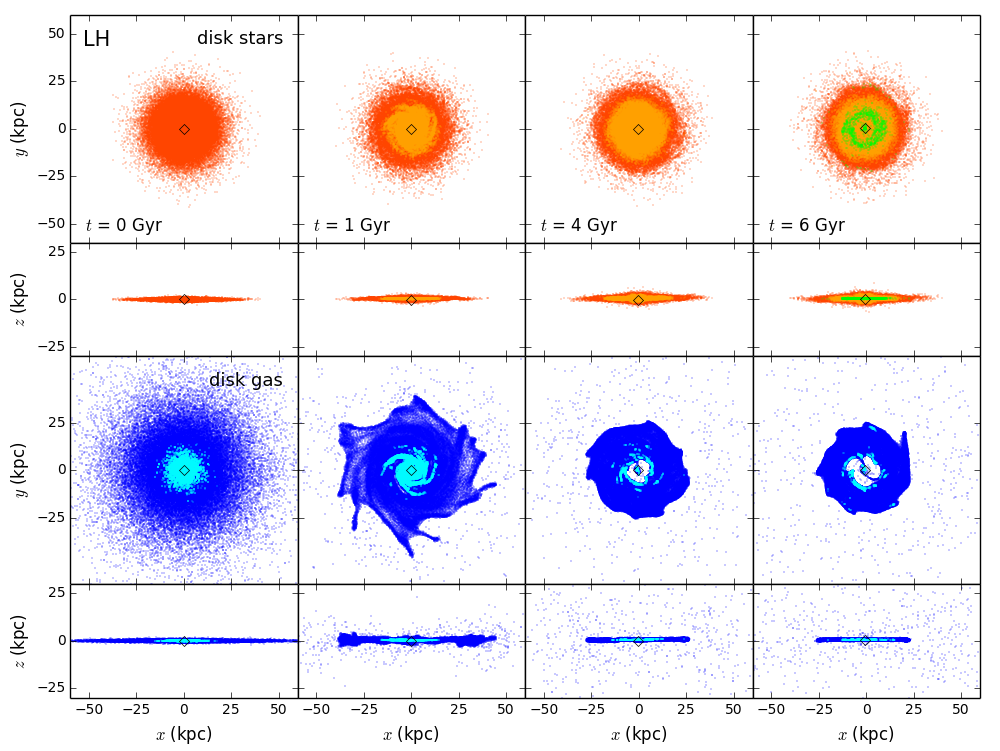}
\caption{
Evolution of the isolated model~LH, which possesses 
both a gas disk and a gas halo 
as well as a stellar disk, a bulge, and a DM halo.
The top two rows display the distribution of 
the star particles on the disk 
seen in the $x$$-$$y$ plane (first row) 
and in the $x$$-$$z$ plane (second row) 
at $t$ = 0, 1, 4, and 6~Gyr 
(from the first to the fourth columns, respectively). 
Orange points represent the original disk star particles 
set from the start of the simulation. 
Light-orange and green points denote stars formed out of the gas 
during the simulation 
before $t$ = 4.75~Gyr and since $t$ = 4.75~Gyr, respectively. 
The bottom two rows show the corresponding views 
of the gas disk. 
Blue points represent the original disk gas particles 
with a zero star formation rate. 
Cyan points are the gas particles having positive values of 
star formation rates. 
}
\end{figure*}

\begin{figure*} 
\centering
\includegraphics[width=16cm]{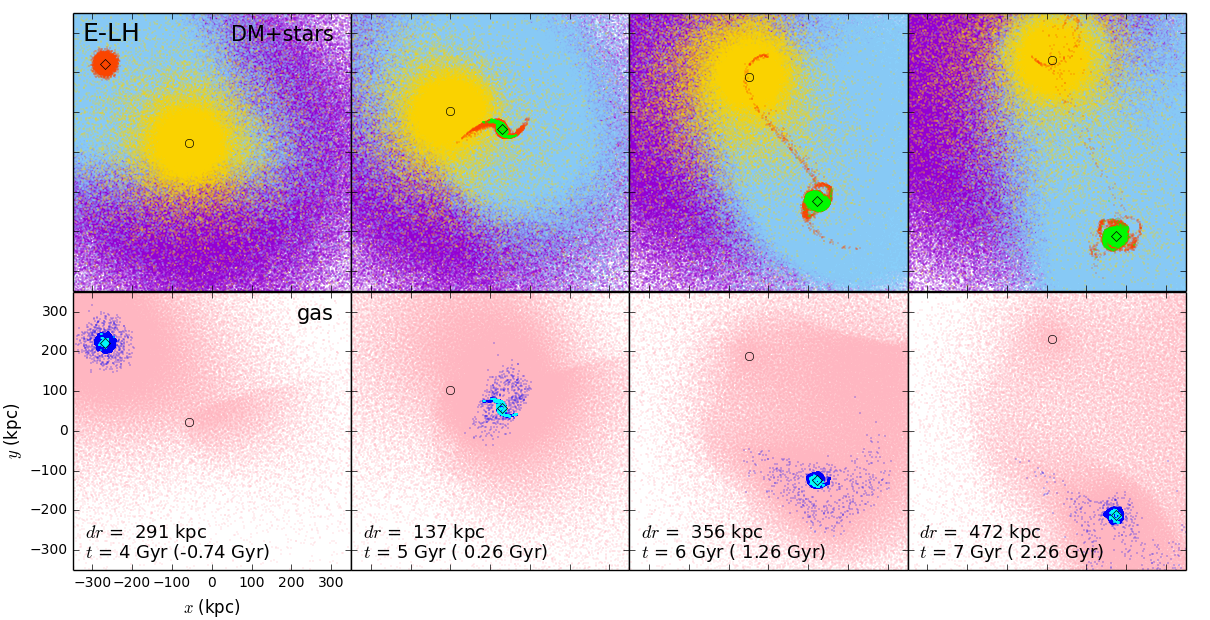}
\caption{The same snapshots as in Figure~2, but from run~E-LH.
The pink points in the bottom panels represent 
the particles originally set as the halo gas of the late-type galaxy~LH. 
}
\end{figure*}

\begin{figure*} 
\centering
\includegraphics[width=14.5cm]{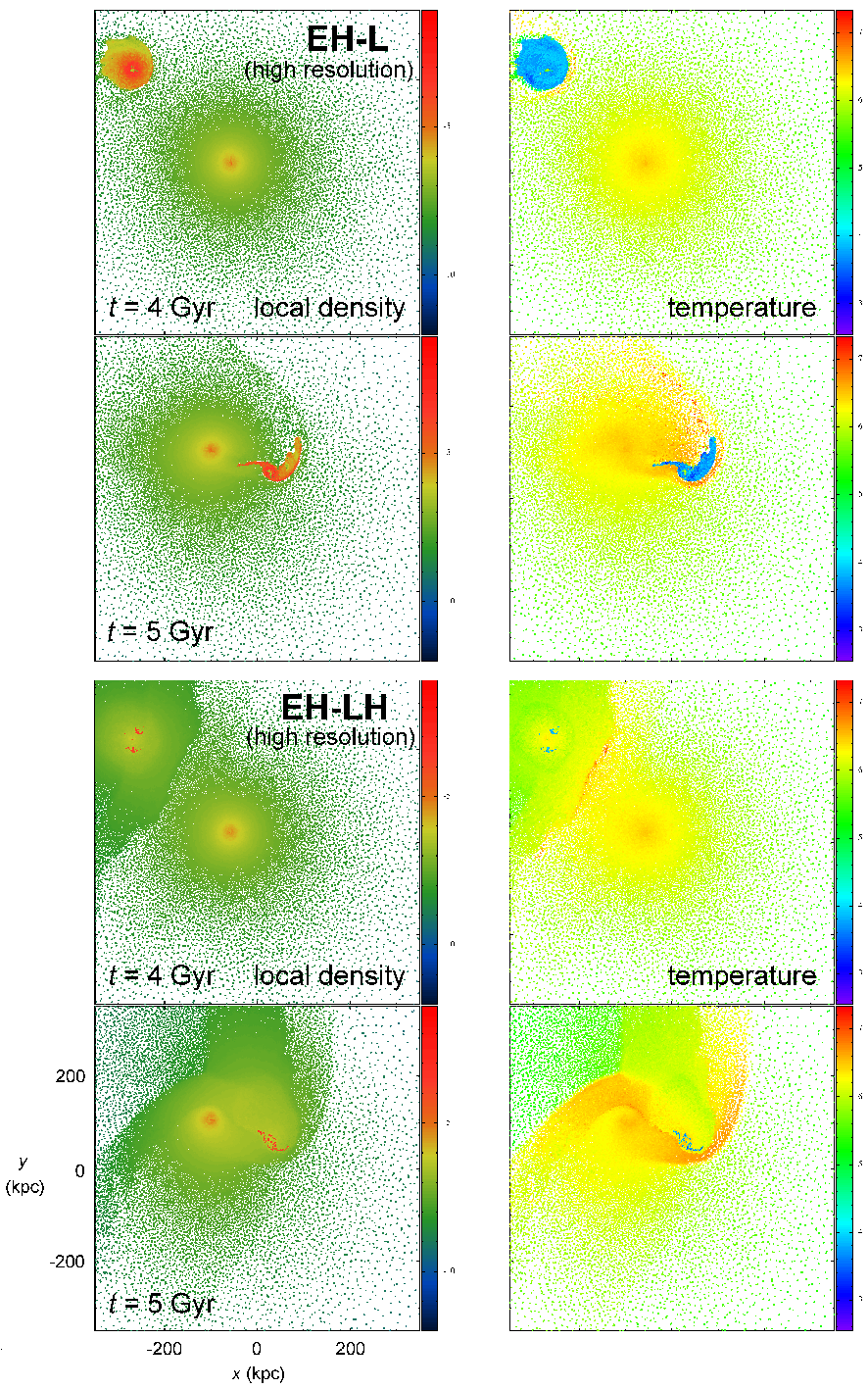}
\caption{
The local density and the temperature 
of the gas particles 
in the high-resolution runs of EH-L (top two rows) 
and EH-LH (bottom two rows) at $t$ = 4 and 5~Gyr. 
As in Figures~7 and 9, 
all gas particles within ${\mid}z{\mid} \le 15$~kpc are displayed, 
with the same ranges of the color bars. 
}
\end{figure*}



\begin{center}
\MakeUppercase{appendix a\\}
\MakeUppercase{Evolution of Model~LH in Isolation}
\end{center}

In order to check the stability of our galaxy models, 
we have run each of our galaxy models in isolation 
for about 7~Gyr.
Figure~12 presents the distribution of the star (top two rows) 
and gas (bottom two rows) particles on the disk 
at $t$ = 0, 1, 4, and 6~Gyr. 
As time passes, some gas particles turn into stars, and 
spiral patterns and a central bar develop in the disk 
(the bar is seen in a more close-up view that 
is not included in this paper).
At later times, 
the gas on the disk becomes dissipated. 
This is primarily because the gas has been continuously consumed 
by star formation, 
and also some of the disk gas has been heated by hot halo gas 
and joined the halo (see the blue points detached from the disk), 
while gas accretion from the halo occurs not very actively. 
The amount of the newly formed stars out of 
the original halo gas particles 
compared to that out of the original disk gas particles is 
less than 1~\%. 
Among the star-forming gas particles (cyan points), 
the percentage of the original halo gas particles 
that cooled down onto the disk is less than 1~\%, as well.  
Compared to model~L without a gas halo, 
the star formation and the gas dissipation in the disk in model~LH 
occur more actively over the course of the simulation 
because of the pressure on the gas disk exerted by the gas halo, 
despite of the gas heating. 
As shown in the figure, the centers of both the star and gas disks 
stay at the origin throughout the simulation. 
We have also checked 
whether the total energy of the model system 
is conserved while it evolves. 
The maximum variation of the total kinetic and potential energies 
is less than 0.9~\% and 0.4~\% of the initial values, respectively.
The maximum change of the total internal energy 
of all of the gas particles 
is less than 7.0~\% of the initial value.

\begin{center}
\MakeUppercase{appendix b\\}
\MakeUppercase{Evolution of Run E-LH}
\end{center}

We have considered the encounter between E and LH, 
where only the less massive LTG has a gas halo,  
for comparison and completeness of our study. 
All of the initial orbital and simulation parameters 
are fixed as in the other three runs.
In run~E-LH, the closest approach occurs at $t$ = 4.74~Gyr 
with a separation between the galaxies of 94.5~kpc. 
Figure~13 shows the snapshots of this run at the same times 
as those in Figure~2.
The distribution of the old disk stars (orange points) 
is almost identical to that in the other three runs. 
At the second snapshot, 
the newly formed stars since the closest approach 
(green points) along the bridge appear less prominently 
compared with run EH-L, 
where the cold disk collides directly with the hot halo of EH.  
In this run, many young stars are found in the disk 
because of the compression by the gas halo of LH, 
as in run~EH-LH. 
The SFR rises as high as in run~EH-LH (see the bottom panel in Figure~10), 
but it increases abruptly right after the closest approach.     
As shown in the bottom panels of Figure~13, 
the halo gas of the LTG (pink points) 
is strongly disturbed by the gravitational pull 
of the more massive ETG. 
The initial disk gas particles 
that have detached from the disk 
follow the motion of the nearby halo gas of the galaxy.

\begin{center}
\MakeUppercase{appendix c\\}
\MakeUppercase{High-resolution Runs}
\end{center}  

We have performed our simulations with a higher resolution as well, 
in order to check particularly whether the shock features 
previously seen in runs EH-L and EH-LH were properly resolved.
In the high-resolution runs, we use 
the LTG models with a four times greater number of particles, 
keeping all of the other model and orbital parameters unchanged. 
(For efficient computing performance, we used 
the same number of particles for the ETG models.) 
Figure~14 presents the local density and the temperature of the 
gas particles near the orbital plane 
(within ${\mid}z{\mid} \le 15$~kpc) from the high-resolution runs. 
As seen in the figure, 
the shocks occurring between the gas disk and the gas halo in run EH-L 
and between the two gas halos in run EH-LH 
are almost identical to those in Figures~7 and 9. 
Because the shocks strongly lead the rest of the evolution, 
the main results of our simulations remain consistent 
in different resolution runs.



\begin{thebibliography}{}

\bibitem[Barnes(2011)]{Barnes2011}
		 Barnes, J. E. \ 2011,
	     Astrophysics Source Code Library, record ascl:1102.027

\bibitem[Barnes \& Hernquist(1992)]{Barnes_Hernquist1992} 
		 Barnes, J. E., \& Hernquist, L.\ 1992,
	     ARA\&A, 30, 705
	     
\bibitem[Barnes \& Hibbard(2009)]{Barnes_Hibbard2009}
		 Barnes, J. E., \& Hibbard, J. E.\ 2009,
	     \aj, 137, 3071
	     
\bibitem[Barnes \& Hut(1986)]{Barnes_Hut1986}
		 Barnes, J., \& Hut, P.\ 1986,
	     \nat, 324, 446	     


\bibitem[Cox(2008)]{Cox+2008}
         Cox, T.~J., Jonsson, P., Somerville, R.~S., 
         Primack, J.~R., \& Dekel, A.\ 2008, 
         \mnras, 384, 386


\bibitem[Genel et al.(2008)]{Genel+2008} 
         Genel, S., Genzel, R., Bouché, N., Sternberg, A., Naab, T., Schreiber, N.~M.~F., 
         Shapiro, K.~L., Tacconi, L.~J., Lutz, D., Cresci, G., Buschkamp, P., 
         Davies, R.~I., \& Hicks, E.~K.~S.\ 2008, 
         \apj, 688, 789
	     
\bibitem[Hernquist(1990)]{Hernquist1990}
		 Hernquist, L.\ 1990,
         \apj, 356, 359
                  
\bibitem[Hwang \& Park(2009)]{Hwang_Park2009}
         Hwang, H.~S., \& Park, C. \ 2009,
         \apj, 700, 791         

\bibitem[Hwang et al.(2013)]{Hwang+2013} 
         Hwang, J.-S., Park, C., \& Choi, J.-H.\ 2013, 
         JKAS, 46, 1 (Paper~I)
         
\bibitem[Katz et al.(1996)]{Katz1996}
		 Katz, N., Weinberg, D. H., \& Hernquist, L.\ 1996,
		 \apjs, 105, 19         

\bibitem[Kennicutt(1998)]{Kennicutt1998}
    	 Kennicutt, R. C. Jr.\ 1998,
		 \apj, 498, 541

\bibitem[L'Huillier et al.(2015)]{L'Huillier+2015}
         L'Huillier, B., Park, C., \& Kim, J.\ 2015, 
         \mnras \, in press (arXiv:1505.00788)


\bibitem[McMillan \& Dehnen(2007)]{McMillan_Dehnen2007}
		 McMillan, P. J., \& Dehnen, W.\ 2007,
		 \mnras, 378, 541


\bibitem[Moster et al.(2011)]{Moster+2011}
		 Moster, B. P., Macci\`o, A. V., Somerville, 
		 R. S., Naab, T. \& Cox, T. J.\ 2011,
		 \mnras, 415, 3750		
		  
\bibitem[Moster et al.(2012)]{Moster+2012}
		 Moster, B. P., Macci\`o, A. V., Somerville, R. S., Naab, T. \& Cox, T. J.\ 2012,
		 \mnras, 423, 2045
	
\bibitem[Naab et al.(2009)]{Naab+2009}
         Naab, T., Johansson, P.~H., \& Ostriker, J.~P.\ 2009, 
         \apj, 699, 178
         
\bibitem[Navarro et al.(1996)]{Navarro+1996}
		 Navarro, J. F., Frenk, C. S., \& White, S. D. M.\ 1996,
		 \apj, 462, 563



\bibitem[Park \& Choi(2009)]{Park_Choi2009}
         Park, C., \& Choi, Y.-Y. \ 2009,
         \apj, 691, 1828

\bibitem[Park et al.(2008)]{Park+2008}	
	     Park, C., Gott, J. R., \& Choi, Y.-Y.\ 2008, 
	     \apj, 674, 784
	     
\bibitem[Park \& Hwang(2009)]{Park_Hwang2009}
         Park, C., \& Hwang, H.~S. \ 2009,
         \apj, 699, 1595	
         
\bibitem[Price D.~J.(2007)]{Price2007}
         Price, D.~J. \ 2007,
         PASA, 24, 159	         
           
	     
\bibitem[Sinha \& Holley-Bockelmann(2012)]{Sinha2012}
         Sinha, M., \& Holley-Bockelmann, K.\ 2012,
         \apj, 751, 17

\bibitem[Smith et al.(2007)]{Smith+2007} 
         Smith, B.~J., Struck, C., Hancock, M., Appleton, P.~N., 
         Charmandaris, V., \& Reach, W~T.\ 2007, 
         \aj, 133, 791

\bibitem[Springel(2005)]{Springel2005}
		 Springel, V.\ 2005,
		 \mnras, 364, 1105
     
\bibitem[Springel et al.(2005)]{Springel+2005}         
		 Springel, V., Di Matteo, T., \& Hernquist, L.\ 2005,
		 \apj, 620, 79
		         

\bibitem[Springel \& Hernquist(2002)]{Springel+2002}
		 Springel, V., \& Hernquist, L.\ 2002,
		 \mnras, 333, 649
		 
\bibitem[Springel \& Hernquist(2003)]{Springel_Hernquist2003}
		 Springel, V., \& Hernquist, L.\ 2003,
		 \mnras, 339, 289		
		 
         
\bibitem[Struck(2006)]{Struck2006} 
         Struck, C.\ 2006,         
         in Astrophysics Update 2, 
         ed. J. W. Mason 
         (Springer Praxis: Chichester), 115 


\bibitem[Toomre \& Toomre(1972)]{Toomre_Toomre1972}
         Toomre, A., \& Toomre, J.\ 1972,
         \apj, 178, 623
         
\bibitem[York et al.(2000)]{York+2000} 
         York, D. G., Adelman, J., Anderson, J. E., et al.\ 2000, 
         \aj, 120, 1579
        

\end{thebibliography}
\end{document}